\documentclass[aps,pre,reprint,superscriptaddress,letterpaper,twocolumn,floatfix]{revtex4-1}

\usepackage{verbatim}
\usepackage{lipsum}
\usepackage{color}
\usepackage{amssymb}
\usepackage{amsmath,mathrsfs}
\usepackage{braket}
\usepackage{bm} % bold math

\usepackage[ruled]{algorithm2e}
\usepackage{algorithmic}
\usepackage{setspace,soul}
\usepackage{graphicx}
\usepackage[dvipsnames]{xcolor}
\usepackage{grffile}
\usepackage{subfigure}

\usepackage[utf8]{inputenc}
\usepackage{geometry}
\newcommand{\dd}{\mathrm{d}}
\usepackage{float}

\newcommand{\pot}{\Phi}
\newcommand{\pathV}{{\mathcal{V}}}

\newcommand{\vsec}{{\bf v}}

\usepackage{xparse}
\ExplSyntaxOn
\NewDocumentCommand{\mref}{m}{\quinn_mref:n {#1}}
\seq_new:N \l_quinn_mref_seq
\cs_new:Npn \quinn_mref:n #1
 {
  \seq_set_split:Nnn \l_quinn_mref_seq { , } { #1 }
  \seq_pop_right:NN \l_quinn_mref_seq \l_tmpa_tl
  ( % print the left parenthesis
  \seq_map_inline:Nn \l_quinn_mref_seq
    { \ref{##1},\nobreakspace } % print the first references
  \exp_args:NV \ref \l_tmpa_tl % print the last or only one
  ) % print the right parenthesis
 }
\ExplSyntaxOff

\usepackage{xr-hyper}
\usepackage{hyperref}

\makeatletter
\newcommand*{\addFileDependency}[1]{% argument=file name and extension
  \typeout{(#1)}
  \@addtofilelist{#1}
  \IfFileExists{#1}{}{\typeout{No file #1.}}
}
\makeatother

\begin{document}

\title{Transition paths in Potts-like energy landscapes: \\
general properties and application to protein sequence models}
\author{Eugenio Mauri, Simona Cocco, R\'emi Monasson}
\email{eugenio.mauri@phys.ens.fr, remi.monasson@phys.ens.fr}
\affiliation{Laboratory of Physics of the Ecole Normale Sup\'{e}rieure, CNRS UMR 8023 and PSL Research, Sorbonne Universit\'e, 24 rue Lhomond, 75231 Paris cedex 05, France}
%\thanks{These two authors contributed equally}

\begin{abstract}
We study transition paths in energy landscapes over multi-categorical Potts configurations using the mean-field approach introduced by Mauri et al., {\em Phys Rev Lett 130, 158402 (2023)}. Paths interpolate between two fixed configurations or are anchored at one extremity only. We characterize the properties of `good' transition paths realizing a trade-off between exploring low-energy regions in the landscape and being not too long, 
such as their entropy or the probability of escape from a region of the landscape.  We unveil the existence of a phase transition separating a regime in which paths are stretched in between their anchors, from another regime, where paths can explore the energy landscape more globally to minimize the energy. This  phase transition is first illustrated and studied in detail on a mathematically tractable Hopfield-Potts toy model, then studied in energy landscapes inferred from protein-sequence data. 
\end{abstract}

\date{\today}

\maketitle

\section{Introduction}

Characterizing transition paths in complex, rugged energy landscapes is a relevant issue in statistical physics and in other fields. In evolutionary biology, for instance, a fundamental problem is to sample mutational paths that, starting from a given protein (described as a given sequence of amino acids) introduce single mutations at each step, %until a known homologous protein is reached, 
in such a way that all the intermediate sequences do not lose their biological activity. Characterizing these paths would be crucial to better understand the navigability of fitness landscapes~\cite{greenbury2022navigability, papkou2023rugged}.
%and how different specific activity in a given protein family may have evolved from ancestral (maybe promiscuous) sequences~\cite{promiscuity}. 
From a statistical mechanics point of view, substantial efforts have been done to characterize how systems dynamically evolve in complex, \emph{e.g.} glassy landscapes to escape from meta-stable local minima and reach lower energy equilibrium configurations. In this context, recent works have focused on   $p$-spin-like energy functions with quenched interactions, generally giving rise to very rugged landscapes \cite{activateddyn_p-spin,ros2021dynamical}. 

Of particular interest is the case of energy functions $E({\bf v})$ defined over Potts-like configurations ${\bf v}=(v_1,v_2, ..., v_N)$, where the variables $v_i$ can take one out of $A$ categorical values \cite{wu1982potts}. Consider a path starting from a configuration ${\bf v}_{\text{start}}$, and exploring $T$ subsequent configurations. The last configuration (extremity) of the path, ${\bf v}_{\text{end}}$, can be free or fixed, depending on the problem of interest. Intermediate configurations along the path can {\em a priori} take any of the $A^N$ possible values, which we refer to as global configuration space below. However, the initial and final configurations define, for each variable $i$, (at most) two categorical values, defining a sub-space with $2^N$ configurations, which we call direct space in the following. The question we address in the present work may be informally phrased as follows: under which conditions are good transition paths naturally living in or close to the direct space rather than in the global one?

\begin{figure}[t]
    \centering
    \includegraphics[width=\linewidth]{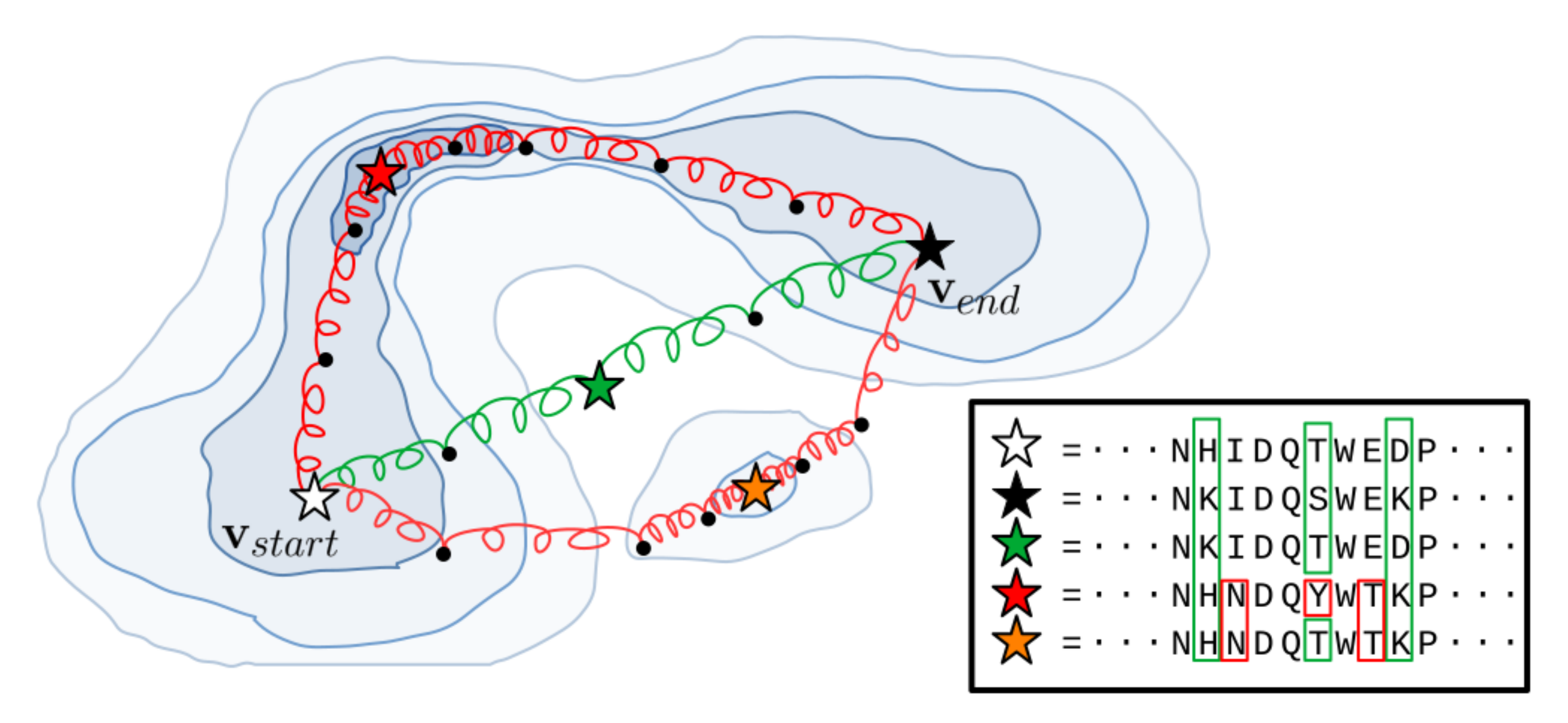}
    \caption{{\bf Mutational paths between two subfamilies in the sequence landscape associated to a protein family.}  Darker blue levels correspond to increasing values of the protein fitness. Paths are either direct (green: each site carries the amino acid present at the same position in the initial or in the final sequence) or global (red: no restriction on amino acids), making possible the exploration of high-fitness regions).}
    \label{fig:sketch}
\end{figure}

This question is of conceptual interest, but has also practical consequences. Consider again the case of mutational paths joining two protein sequences, see Fig.~\ref{fig:sketch}. Due to the huge number of possible paths in the global space ($A=20$), mutagenesis experiments generally restrict to direct paths~\cite{Poelwijk2019Sep}. However, constraining paths to be direct may preclude the discovery of much better global paths, involving mutations and their reversions and reaching more favorable regions in the sequence space (Fig.~\ref{fig:sketch}). Mutational experiments have demonstrated that exploring the fitness landscape beyond the direct space can enhance adaptation \cite{Wu_elife_indirect}. Additionally, the existence of such beneficial `global' mutations could provide valuable insights into the properties of the fitness landscape, {\em e.g.} the presence of high fitness regions responsible for the deviation of the paths from the direct subspace. 

Whether paths remain direct or explore the global space will depend on their length, on their `elastic' properties (defined by the mutation process), as well as on the nature of the energy (minus fitness) landscape. In particular, fitness landscapes can be complex and with many good regions (surrounding local maxima) that attract the path outside the direct space. In this article, we introduce a minimal landscape model, corresponding to a Hopfield-Potts model with $P$ patterns, defining a rank $P$ pairwise coupling matrix between the $v_i$'s. The energy of a path is then defined as the sum of the Hopfield-Potts energies of the intermediate configurations, and of elastic contributions measuring the dissimilarities between successive configurations (Fig.~\ref{fig:sketch}). When $N$ is sent to infinity while keeping $P$ finite, transitions paths in this landscape can be analytically studied using the mean-field framework introduced in \cite{mauri2023}. Two sets of time-dependent order parameters, where the time $t$ denotes the coordinate along the path are needed in the mean-field theory: (1) the average projections $m_t^\mu$ of configurations along the patterns $\mu=1,...,P$; (2) the overlaps $q_t$ between successive configurations ${\bf v}_t$ and ${\bf v}_{t+1}$ along the path. We explain how the mean-field theory allows for a detailed study of the statistical properties of transition paths, such as their entropy, the escape probabilities from a local minima, and their direct vs. global nature. In particular, we show that, depending on the stiffness coefficient of the elastic term acting on $q_t$ two regimes can be encountered. Paths under high tension are likely to remain confined within the direct space. For low tension, paths are likely to explore the global path to minimize their energy. The nature of the phase transition, such as the critical tension and the time-behaviour of the order parameters are analytically unveiled. We also compute the entropy of transition paths interpolating between the anchoring (initial and final) configurations.

Importantly, the mean-field formalism can be transferred to restricted Boltzmann machines (RBM) trained from natural protein sequence data \cite{fischer2012introduction,tubiana_learning_2019}. While protein-sequence landscapes are {\em a priori} unknown a vast use of data-driven models managed, over the past years, to capture the relation between protein sequences and their functionalities. Unsupervised machine-learning approaches such Boltzmann machines or Variational AutoEncoders could be trained from homologous sequence data and used to score the sequences, hence defining an empirical energy, and were in particular shown to be  generative, \emph{i.e.} they could be used to design novel proteins with functionalities comparable to natural proteins~\cite{Russ2020Jul,Hawkins2021}. In this context, RBM can be seen as a natural extension of Hopfield-Potts models, where the weights connecting the visible (sequence) and hidden (representation) layers play the role of patterns, and the energy is not necessarily quadratic in the projections $m^\mu$. We hereafter apply RBM to sequence data coming from {\em in silico}  and real  proteins, and show that direct-to-global phase transitions are found in transition paths built from such data-driven models.

This paper is organized as follows. Section~\ref{overview} provides the main definitions and an overview of the basic properties of transition paths in Hopfield-Potts landscapes. In Section~\ref{interacting_seqs} we recall the transition-path framework introduced in \cite{mauri2023}, in particular the expression of the mean-field free-energy as a function of the order parameters $\{m_t^\mu,q_t\}$ for a generic Hopfield-Potts energy, and study in detail the direct-to-global phase transition in the minimal case of $P=2$ non-orthogonal patterns. In Section~\ref{sec:inferred} we apply our mean-field approach to energy landscapes inferred from  {\em in silico} lattice-protein models \cite{Jacquin_LP_2016} to benchmark our approach, and from natural protein sequence data associated to the WW domain, a short protein domain involved in signalling~\cite{tubiana_learning_2019}. Conclusive remarks can be found in Section~\ref{sec:conclusion}.

%In~\cite{mauri2023} we have already carried out our approach on protein sequence data. In particular, we have introduced a discrete-steps transition path sampling algorithm, and the Mean-field theory under two different elastic potentials characterizing the rigidity of a transition path (Fig.~\ref{fig:sketch}). The first  enforcing a constant  number of mutations at each  evolutionary step, the second inspired by Kimura theory~\cite{kimuraneutral} enforcing a constant evolutionary rate for each site on the sequence. We have moreover characterized the optimal transition paths, discussed them from a evolutionary-biology point of view, and compared the different evolutionary dynamics. We will hereafter refer to these two possible choices as Cont and Evo scenario.

\section{Definitions and overview of the results}
\label{overview}

 %From an evolutionary perspective, the objective is to reconstruct the most likely evolutionary trajectories between two known sequences within the framework of rare event sampling. This can facilitate the development of novel phylogenetic reconstruction algorithms based on complex models of evolution, where the fitness landscape driving selection exhibits high epistasis, \emph{i.e.}, the effect of each mutation strongly depends on the entire sequence. 

\subsection{Mutational paths over configuration space}

We consider an energy landscape  $E_{\text{model}}(\mathbf{v})$ over $N$-dimensional Potts configurations $\mathbf{v}$, see Fig.~\ref{fig:sketch}.  $E_\text{model}$ can be either derived from first principles, or inferred from some available data using machine-learning methods. Following \cite{mauri2023}, 
%we model the probability of a path with extremities ${\bf v}_{\text{start}}$, ${\bf v}_{\text{end}}$. 
we associate to each path $\pathV =\{{\bf v}_{\text{start}},{\bf v}_1,{\bf v}_2,..., {\bf v}_{T-1},{\bf v}_{\text{end}}\}$  an energy $\mathcal{E}(\pathV)$. This energy is the sum of the energies of the intermediate configurations along the path, and of elastic contributions decreasing with the similarities between pairs of successive configurations. We denote by $\Phi$ the elastic potential.  The energy of a path (divided by $N$) is then  
\begin{multline}
   \mathcal{E}(\pathV ) = \frac 1N \sum_{t=1}^{T-1} E_{\text{model}}(\vsec_t) +  \pot(q({\bf v}_{\text{start}} , \vsec_1)) \\+ \sum_{t=1}^{T-2} \pot(q(\vsec_t, \vsec_{t+1})) + \pot(q({\bf v}_{T-1} , \vsec_{\text{end}}))\,,
    \label{eq:path_weight_en}
\end{multline}
where the overlap $q(\vsec_{t},\vsec_{t+1}) = \frac{1}{N} \sum_i \delta_{v_{i,t}, v_{i,t+1}}$ measures the similarity between adjacent sequences. 

The probability of the path is then defined as the Boltzmann distribution 
\begin{equation}
    \mathcal{P}[\pathV ] =\frac 1{Z_{\text{path}}}   \; e^{-\beta \, N\, \mathcal{E}(\pathV ) }\ ,
        \label{eq:path_weight}
\end{equation}
where $\beta$ is an inverse temperature and $Z_{\text{path}}$ ensures normalization.
This distribution promotes paths, where intermediate configurations have low energies $E_\text{model}$,  and are not far away from each other in order to guarantee smoothness in the interpolation. A key role is played by the potential $\pot$, which controls the elastic properties of the path. In~\cite{mauri2023}, we considered two choices for $\pot$, corresponding to distinct scenarios for the mutational dynamics. The first one, denoted by Cont, makes sure that any two contiguous configurations along the path, $\mathbf{v}_t$ and $\mathbf{v}_{t+1}$, differ by a bounded (and small compared to $N$) number of sites. The second choice for $\pot$ is inspired by Kimura's theory of neutral evolution~\cite{kimuraneutral} and hereafter called Evo. It enforces a constant mutation rate for each variable, see below. 

In the Cont scenario, we aim to build paths that continuously interpolate between the two target configurations as $T$ growths. Hence, we choose $\pot$ in order to avoid small overlaps $q$ between adjacent sequences, which would signal large jumps along the path. 
In practice, we set
\begin{equation}\label{eq:potcont}
    \pot_{\text{Cont}}(q) = \frac{1}{T^2 |q-q_c|} =  \frac{1}{T^2 \left|q-1+\frac{\gamma}T\right|}\,,
\end{equation}
where the $1/T^2$ scaling in the potential guarantees the existence of continuous solution in the large-$T$ limit as shown in Section~\ref{HP_model}; Other choices of potentials with hard-wall constraints give similar results. The parameter $\gamma$ controls the elasticity of the path. Its minimal value is $D/N$, where $D$ is the Hamming distance between the extremities $\vsec_{\text{start}}$ and $\vsec_{\text{end}}$. Larger values of $\gamma$ will authorize more flexible paths. 

In the Evo scenario, the potential $\pot$ is chosen to emulate neutral evolution with a certain mutation rate $\mu$, and is given by
\begin{equation}
    \Phi_\text{Evo} (q) =( 1-q) \ln \left( 1 + \frac{A}{e^{\,\mu A/(A-1)}-1}\right) \ .
\end{equation}
In this setting, paths can be seen as alternating steps of random mutations (starting from $\vsec_{\text{start}}$) and of selection, parametrized by, respectively, the mutation rate $\mu$ and the effective `log. fitness' $-E_{\text{model}}$. The transition path is conditioned to end in  $\vsec_{\text{end}}$. In standard evolutionary dynamics, paths are not constrained by their final configuration, but only by their initial one. Such paths are anchored at one extremity only. However, if the configuration (genome) of an organism is observed after some evolutionary time, it is legitimate to ask about the  distribution of putative paths followed by the organism that interpolate between this 'final' and the known initial configurations. Asa result of this conditioning, the transitions paths are now anchored at both extremities.

\subsection{Minimal Hopfield-Potts model for transition paths}
\label{sec:mhp78}

We now introduce a minimal setting, where the properties of transition paths can be analytically characterized. 

\subsubsection{The Hopfield-Model landscape}

We first define the energy landscape for Potts configurations.
We consider a Hopfield model for categorical data, hereafter referred to as Hopfield-Potts. There are $A\ge3$ states per site (called $a$, $b$ and $c$ and so on). The energy of our Minimal Hopfield-Potts (MHP) model reads
\begin{equation}%\label{ener_HP}
    E_{\text{MHP}}(\mathbf{v})= - \frac{J}{2N}\sum_{i,j} \big( w_{1i}(v_i)w_{1j}(v_j) + w_{2i}(v_i)w_{2j}(v_j) \big)\, ,
    \label{eq:HP_energy_func}
\end{equation}
where the two patterns $\mathbf{w}$ are constructed as follows: 
\begin{align}
    w_{1i}(v_i) = \delta_{v_{i},a} +  \omega\, \delta_{v_{i},c} \ , \nonumber \\
    w_{2i}(v_i) = \delta_{v_{i},b} +  \omega \, \delta_{v_{i},c}\ , 
    \label{eq:patterns_HP}
\end{align}
and $\omega$ is a positive parameter that controls how much the two patterns overlap. The coupling strength $J$ is supposed to be large, but its precise value does not affect the qualitative description below. The energy of a configuration $\mathbf{v}$ is a quadratic function of its two projections along the patterns, denoted as $m^{\mu}(\mathbf{v}) = \frac 1N \sum_i w_{i\mu}(v_i)$ ($\mu = 1, 2$):
\begin{equation}
    E_{\text{MHP}}(\mathbf{v})= - \frac{J}{2}\, N\, \bigg[ \big( m^{1}(\mathbf{v})\big )^2 +  \big( m^{2}(\mathbf{v})\big )^2 \bigg]\, ,
    \label{eq:HP_energy_func2}
\end{equation}
The MHP model is therefore intrinsically of mean-field nature, and can be easily solved in the large-$N$ limit.

%The HP model is designed to store $M$ distinct patterns $\textbf{w}$ and compute the energy of a configuration $\mathbf{v}$ through evaluating i, hereafter referred to as projection. This order parameter is similar to Mattis projection, and the square values of $m_{\mu}$ are summed. In our simplified model, we set $M = 2$ and $A\ge3$ states per site (called $a$, $b$ and $c$ and so on). We will consider the thermodynamic limit $N\to \infty$. The 
%attractiveness of the $c$ direction in the Potts space.

\begin{figure*}
    \centering
    \includegraphics[width=\linewidth]{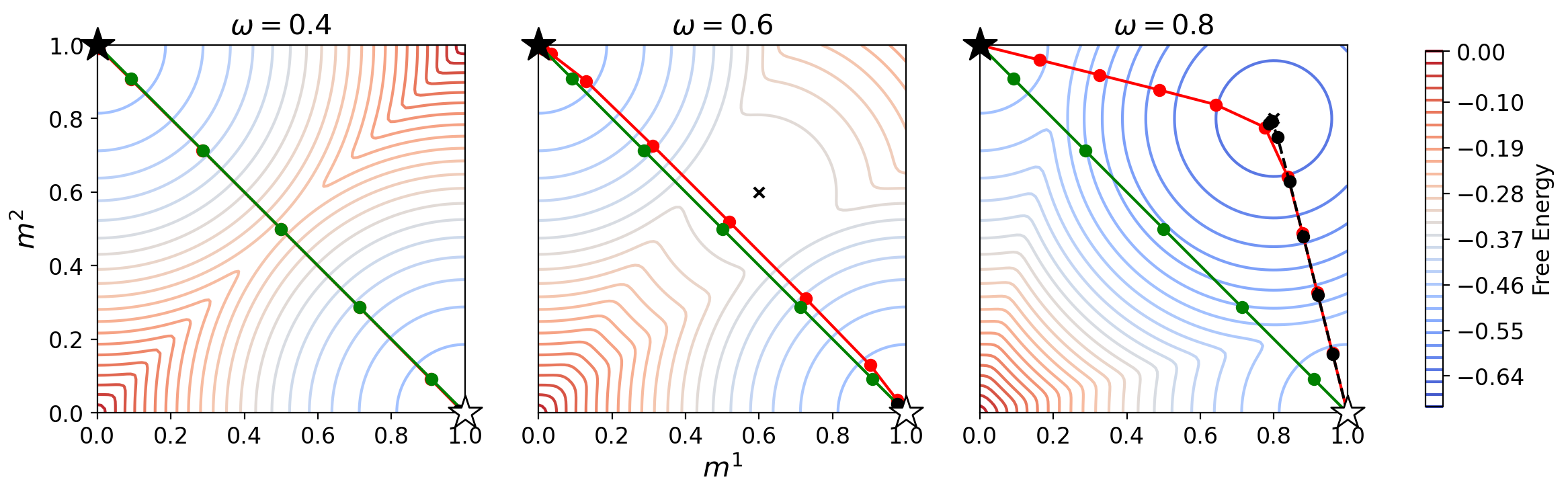}
    \caption{{\bf Sketch of the direct-to-global phase transition in the MHP model}. Green paths correspond to path constrained in the direct space, while red are global (free to explore any configuration). Stars represent the initial and final configurations, see corners of the free-energy landscape. Each plot represents the HP energy landscape for a fixed value of $\omega$, showing the crossover between direct and global transition paths as the symmetric minimum of the free energy becomes more and more attractive, {\em i.e.} as the overlap between the patterns $\omega$ increases and as the length of the path increases. Black dashed lines correspond to paths computed without fixing the end, showing a transition between paths staying close to initial minimum (\emph{i.e.} the white star) and paths that jump into the intermediate minimum when this becomes stable for higher values of $\omega$. Here the parameters $\beta$ and $\gamma$ appearing in Eqs.~(\ref{eq:path_weight}) and (\ref{eq:potcont}) are equal to, respectively, $6$ and $3$. The length of all paths defined in Eq.~\eqref{eq:path_weight_en} is set to $T=10$, but only points that are different from the endpoints (\emph{i.e.} white and black stars) are shown.}
    \label{fig:sketch_HP_over}
\end{figure*}

A sketch of the free-energy of the MHP model in the $(m^1,m^2)$ plane is shown in Figure~\ref{fig:sketch_HP_over}. Depending on the value of $\omega$ two cases must be be distinguished:
\begin{itemize}
\item For $\omega<\frac 12$, the only minima of the free energy are $(m^1,m^2) = (m^*,0)$ and $(0,m^*)$, with $m^*\simeq 1$. As a consequence, the only configuration with non--negligible probabilities are the two patterns themselves. 
\item For $\omega>\frac 12$, a new local minimum will appear at $(m^1,m^2)=(\omega, \omega)$, which we refer to as symmetric minimum later. This local minimum becomes global when $\omega >\frac 1{\sqrt 2} $. Therefore, the energy landscape includes a region, far away from the pattern-associated configuration, which is energetically favorable.
\end{itemize}

\subsubsection{Transition paths anchored at both extremities}

In this landscape, we will consider paths of configurations anchored at both extremities, {\em i.e.} such that $\mathbf{v}_{\text{start}}=\{a,a,a,...,a\}$ and  $\mathbf{v}_{\text{end}}=\{b,b,b,...,b\}$.  For the sake of simplicity, we will restrict ourselves to the Cont potential, see Eq.(\ref{eq:potcont}). As the Hamming distance between the two edges of the path is equal to $D=N$, the flexibility parameter $\gamma$ must be larger than 1.

The properties of the mutational paths associated to this energy landscape can be analytically characterized. The mean-field theory associated to paths is more sophisticated than for single configurations. Explicit expressions can nevertheless be derived for the average projections $m^1_t,m^2_t$ of intermediate configurations $\mathbf{v}_t$ and for the average overlap  $q_t$ between successive sequences $\mathbf{v}_t,\mathbf{v}_{t+1}$;   Detailed calculations and results are reported in Section~\ref{interacting_seqs}. 
Briefly speaking, we find that, see Fig.~\ref{fig:sketch_HP_over}:
\begin{itemize}
\item For $\omega<\frac 12$, the optimal path connecting the starting and ending configurations is direct. Due to the absence of favorable regions in the landscape outside the neighborhoods of the anchors paths have no incentive to explore the landscape: they directly interpolate between $\mathbf{v}_{\text{start}}$ and $\mathbf{v}_{\text{end}}$ to minimize their elastic energy.
\item For $\omega>\frac12$, the symmetric minimum attracts mutational paths and make them leave the direct space. If paths are sufficiently long and flexible they are deviated by this minimum, and explore the global configuration space.
\end{itemize}
While the precise locus of the paths are specific to the MHP model, the coincidence of the onset of the transition with the existence of favorable regions in the configurations space is a general phenomenon. The nature of optimal transition paths is therefore intimately related to the structure of the energy landscape.

\subsubsection{Transition paths anchored at one extremity}

We consider the case in which $\mathbf{v}_{\text{end}}$ is not fixed. In this scenario, this final configuration is very likely, for large $N$,  to lie in one of the global minima of the free energy.  If the starting configuration is attached to another minimum, then paths  can explore the space in its diversity, see Fig.~\ref{fig:sketch_HP_over}(right) for an illustration. 

Our mean-field theory can be adapted to the case of paths anchored at one extremity,  and allows us to estimate the time, i.e., the minimal length necessary for a path to escape some region $\mathcal{R}$ of the configuration space. In practice, a region is defined as the local minimum of the mean-field free energy containing $\mathbf{v}_{\text{start}}$. We define the probability of paths of length $T$ to stay in $\mathcal{R}$ through the ratio of the statistical weight, defined in Eq.~\eqref{eq:path_weight}, of all the paths constrained to end in $\mathcal{R}$ and the weight associated to unconstrained paths, {\em i.e.} free to wander in the configuration space. This probability reads
\begin{align}
   P_{\text{stay}}(\mathcal{R}|T) &=  \frac{\displaystyle{\sum_{\pathV:\mathbf{v}_{\text{end}}\in \mathcal{R}}e^{-N\mathcal{E}(\pathV)}}}{
   \displaystyle{\sum_\pathV e^{-N\mathcal{E}(\pathV)}}}
   \nonumber \\ &
    \underset{{\scriptscriptstyle N\gg1}}{\sim} \frac{e^{-Nf_{\text{path}}^{\text{const.}}(\mathbf{v}_{start},\mathcal{R}|T)}}{e^{-Nf_{\text{path}}^{\text{unconst.}}(\mathbf{v}_{start}|T)}}\,,
    \label{eq:escape_prob_HP}
\end{align}
where $f_{\text{path}}^{\text{const.}}$ and $f_{\text{path}}^{\text{unconst.}}$ are the free energies associated to, respectively,  constrained and unconstrained paths. The calculation of these free energies is reported  in Section~\ref{sec:MF} and~\ref{sec:escape_HP}. For the MHP model, we observe that, above a certain crossover value of $\omega$ (depending on $T$), the path is very likely to escape from the local minimum at $(m^1,m^2)=(m^*,0)$ and to end up in $(\omega, \omega)$, see Fig.~\ref{fig:sketch_HP_over}.

\subsection{Transition paths with Restricted Boltzmann Machines inferred from protein sequence data}

Our mean-field approach for transition paths can be applied to more complex Hopfield-Potts energies than Eq.~\eqref{eq:HP_energy_func2}, {\em i.e.} with more than two patterns, and/or with non-quadratic dependence on the projections $m^\mu$. This is the case for the so-called Restricted Boltzmann Machines (RBM), class of unsupervised architectures that can be trained from data.

\subsubsection{Restricted Boltzmann Machines and landscape inference}

Generally speaking, unsupervised machine learning aims to infer an energy landscape through the inference of a probabilistic model $P_{\text{model}}({\bf v})$ from data configurations, ${\bf v}_b, b=1,...,B$.  We consider RBM, a bipartite neural network, in which data configurations $\bf v$ are carried by a $N$-dimensional layer of visible neurons, and representations $\bf h$ of these data are extracted by a $M$-dimensional layer of real-valued hidden (latent) units. The two layers interact through the weights $w$. The joint probability distribution of visible and hidden configurations is given by, up to a normalization constant,
\begin{multline}
    P_{\text{RBM}} (\mathbf{v},\mathbf{h})\propto \exp \left( \sum_i g_i(v_i) + \right. \\ \left.+\sum_{\mu} h_\mu I_\mu({\bf v}) - \sum_\mu \mathcal{U}_\mu(h_\mu) \right)\ ,
\label{eq:rbm}
\end{multline}
where  $I_\mu({\bf v}) = \sum_i w_{i,\mu}(v_i)$ is the input to hidden unit $\mu$. The $g_i$'s and $\mathcal{U}_\mu$'s are local potentials acting on, respectively, visible and hidden units. Note that the weight $w_{i\mu}(v_i)$ between visible unit $i$ and hidden unit $\mu$ depend on the category $v_i$ of the visible unit.  The hidden potentials $\mathcal{U}_\mu$ are chosen among the class of double Rectified Linear Units (dReLU):
\begin{equation}
    \mathcal{U}_\mu(h) = \frac{1}{2}\gamma_{\mu,+}h_+^2 + \frac{1}{2}\gamma_{\mu,-}h_-^2 + \theta_{\mu,+}h_+ + \theta_{\mu,-}h_-\,, 
    \label{dRelu_rbm}
\end{equation}
where $h_+ = \max(h,0)$ and $h_- = \min(h,0)$ \cite{tubiana_learning_2019}. All the parameters of the model (weights and local potentials) are learned by maximizing  $\prod_{b=1}^B P_{\text{RBM}}({\bf v}_b)$ using Persistent Contrastive Divergence~\cite{Tieleman2008Jul}; regularization over model parameters can also be enforced. Here, $P_{\text{RBM}}({\bf v}) = \int d{\bf h}\, P_{\text{RBM}}({\bf v},{\bf h})$ is the marginal distribution for configurations. As a result the RBM energy is
\begin{eqnarray}\label{eq:tyujk}
    E_{\text{RBM}}(\mathbf{v}) &=& - \log P_{\text{RBM}}({\bf v})\nonumber \\
    &=&- \sum_{i=1}^N g_i(v_i) - \sum _{\mu=1}^M \Gamma_\mu\big( I_\mu (\mathbf{v})\big) \ ,
\end{eqnarray}
where $\Gamma_\mu(I)=\log \int dh \, e^{-{\cal U}_\mu(h) +h I}$ and irrelevant additive constants have been omitted. 

The expression of $E_\text{RBM}$ above shows that RBM are a generalized class of Hopfield-Potts models. In addition to local potentials acting on the visible units ($g_i$), the energy depends on the configuration $\mathbf{v}$ through the inputs $I_\mu (\mathbf{v})$ only. These inputs play the same role as the projections $m^\mu(\mathbf{v})$ in the Hopfield-Potts framework; both quantities are simply related through $ m^\mu (\mathbf{v})= \frac 1N I_\mu (\mathbf{v})$. We stress that the dependence of the energy upon the inputs is generally non quadratic. Standard Hopfield-Potts models are recovered for ${\cal U}(h)\propto h^2$, implying $\Gamma(I)\propto I^2$. The number of patterns is, in the context of RBM, equal to the number $M$ of hidden units. In pratice, $M$ is an hyper-parameter which is fixed during learning through cross-validation procedures. The Hopfield-Potts nature of RBM allows us to straightforwardly extend our mean-field approach to these data-driven models, see Section~\ref{sec:mfrbm}.

\subsubsection{Applications to proteins}

We apply in Sec.\ref{sec:inferred} our analytical mean-field Hopfield-Potts framework to the RBM energy landscapes inferred from sequence data of real  and synthetic protein families. All necessary information about training and sequence data can be found in Sections~\ref{sec:prot1w} \& \ref{sec:prot2w} and in \cite{mauri2023}.

We observe the same kind of direct-to-global transition as the one discussed for the MHP model above.  Moreover we compute, for the WW domain, the entropy of paths as a function of their length for both Cont and Evo potentials, with or without fixed end extremity, as well as the probability of staying in the initial region of the energy landscape.  An outcome of this work, of pratical relevance to mutagenesis experiments, is the prediction of the sites $i$ and amino acids $v_i$, where mutations outside the direct space are expected to be highly beneficial. These predictions could be used to propose and test new mutations along transition paths, and offer a controled way to explore the sequence space beyond the amino acids present in the initial and final proteins. In our toy model of lattice proteins such reversed mutations are essential to stabilize the protein when paths join two functionally distinct regions, and show switching from one specificity to another.

\section{Mean-field theory and direct-to-global transition for the Minimal Hopfield-Potts model}
\label{interacting_seqs}

In this section we describe the mean-field theory treatment of paths in the MHP landscape following \cite{mauri2023}, solve the corresponding self-consistent equations for the order parameters $\{m^1_t,m^2_t,q_t\}$ along the path, and then characterize the nature of the transition. For the sake of generality, expressions are written for a generic number $M$ of patterns $w_{i\mu}(v)$, and then applied to the case of the $M=2$ patterns in Eq.~\eqref{eq:patterns_HP}.

\subsection{Mean-field theory of transition paths}
\label{sec:MF}

%We consider the energy \eqref{eq:HP_energy_func} for the MHP model with $M=2$ patterns can be straightforwardly extended to the case of $M$ patterns, with components $w_{i\mu}(v_i)$. The probability distribution, $P_{\text{model}}(\mathbf{v}) \propto \exp [- E(\mathbf{v})]$, can equivalently be defined as the marginal distribution of a model, whose energy is a function of the visible configuration $\bf v$ and the hidden real-valued vector $\bf h$:
%\begin{equation}
%    \label{eq:HP_with_hidden}
%    E_{\text{full}}(\vsec, \textbf{h}) = - \sum_{i,\mu} w_{i\mu}(v_i)h_\mu + \frac N2 \sum_\mu h_\mu^2.
%\end{equation}
%This model is equivalent to a RBM with $M$ hidden units and quadratic local potentials:
%\begin{equation}
%    E(\mathbf{v},\mathbf{h}) = - \sum_{i,\mu} w_{i\mu}(v_i)h_\mu + \frac{N}{2}\sum_\mu h_\mu^2\,.
%\end{equation}

The partition function $Z_{\text{path}}$ defined in Eq.~\eqref{eq:path_weight} with the energy function in Eq.~\ref{eq:HP_energy_func} can be expressed 
%computed in the $N\to\infty$ limit using mean-field theory. To this aim we define the average projections of $\vsec$ along the patterns $w_\mu$ at each step of the path as $m_t^\mu=\frac 1N \sum_i \langle w_{i\mu}(v_{it}) \rangle$, while the average overlap between adjacent sequences is given by $q_t= \langle q(\mathbf{v}_t,\mathbf{v}_{t+1}) \rangle$. Here,  the average is computed over the path distribution  in Eq.~\eqref{eq:path_weight}, $\langle \cdot \rangle = \sum_\pathV (\cdot) \exp(-\beta \mathcal{E}(\pathV))/Z_{\text{path}}(\beta)$, where $\beta$ is an inverse temperature. Using
as a integral over the projections $\mathbf{m} = \{m_t^\mu\}$  of intermediate configurations on the patterns and over the overlaps $\mathbf{q} = \{q_t\}$ between successive configurations:
\begin{multline}
    Z_{\text{path}}(\beta) = \int \dd \mathbf{m}\,\dd \mathbf{q} \exp\left[ \frac{N\beta}{2}\sum_{\mu,t}{(m_t^\mu)^2} \right. \\ \left. - N\beta\sum_t\pot(q_t) + N\, \mathcal{S}(\mathbf{m}, \mathbf{q})\right]\,,
    \label{eq:part_func_HP_path}
\end{multline}
where we have defined the entropy as
\begin{multline}
    \label{eq:entropy_HP_path}
   \mathcal{S}(\mathbf{m}, \mathbf{q}) = \frac 1N \log \sum_\pathV \prod_{\mu,t} \delta\left(\frac 1N \sum_i  w_{i\mu}(v_{i,t})- m_t^\mu\right)\\ \times \prod_{t}\delta\left(\frac 1N \sum_{i=1}^N \delta_{v_{i,t},v_{i,t+1}} -q_t\right)\,.
\end{multline}
Using integral representations of the Dirac $\delta$'s, we may express the entropy as an integral over the auxiliary variables $\hat{\mathbf{m}} = \{\hat{m}_t^\mu\}$ and $\hat{\mathbf{q}} = \{\hat{q}_t\}$:
\begin{multline}
    \mathcal{S}(\mathbf{m}, \mathbf{q}) = \frac 1N \log \int   \frac{\dd \hat{\mathbf{m}} \dd \hat{\mathbf{q}} }{(2\pi/N)^2}\exp\left[ -N\mathbf{m}\cdot\hat{\mathbf{m}} -N\mathbf{q}\cdot\hat{\mathbf{q}} \right]\\ \times \sum_\pathV  \prod_i\exp
   \left[ \sum_{\mu,t}\hat{m}_t^\mu w_{i\mu}(v_{i,t}) + \sum_{t}\hat{q}_t\,\delta_{v_{i,t},v_{i,t+1}} \right]\,.
\end{multline}
In the large--$N$ limit we obtain
%We can now write $\sum_\pathV \prod_i \cdots = \prod_i \sum_{\{v_{it}\}_t} \cdots$ and further develop the previous equation as
%\begin{multline}
%    \mathcal{S}(\mathbf{m}, \mathbf{q}) = \frac 1N \log \int \dd \hat{\mathbf{m}} \dd \hat{\mathbf{q}} \exp\Bigg[ -N\mathbf{m}\cdot\hat{\mathbf{m}} -N\mathbf{q}\cdot\hat{\mathbf{q}} + \\ +   \sum_i \log Z_i^{\text{1D}} \Bigg]\,,
%\end{multline}
\begin{equation}
     \mathcal{S}(\mathbf{m}, \mathbf{q}) = \min_{\hat{\mathbf{m}} , \hat{\mathbf{q}}} \left[ -\mathbf{m}\cdot\hat{\mathbf{m}} -\mathbf{q}\cdot\hat{\mathbf{q}} + \frac 1N \sum_i \log Z_i^{\text{1D}} (\hat{\mathbf{m}},\hat{\mathbf{q}})\right]\, ,
     \label{eq:entropy}
\end{equation}
where
\begin{equation}
    Z_i^{\text{1D}} = \sum_{\{v_{t}\}} \exp\left[ \sum_{\mu,t}\hat{m}_t^\mu w_{i\mu}(v_{t}) + \sum_{t}\hat{q}_t\,\delta_{v_{t},v_{t+1}} \right]\,
    \label{eq:logZ_1d}
\end{equation}
is the partition function of a 1D-Potts models with nearest-neighbour interactions. We note that in the case of paths with both ends fixed the starting and final element of this sum are fixed, while for paths with free ends we also sum over the last element $v_T$.
%The integral in $\mathcal{S}$ is dominated by the terms that maximise the argument in the exponential. Hence, we write
At the saddle-point, the auxiliary variables  fulfill the following set of coupled implicit equations:
\begin{align}
    m_t^\mu &= \frac{1}{N} \sum_i \frac{\partial \log Z_i^{\text{1D}}}{\partial \hat{m}_t^\mu}(\hat{\mathbf{m}},\hat{\mathbf{q}})\ ,\nonumber \\
    q_t &= \frac{1}{N} \sum_i \frac{\partial \log Z_i^{\text{1D}}}{\partial \hat{q}_t}(\hat{\mathbf{m}},\hat{\mathbf{q}})\ .
    \label{eq:der_logZ_1d}
\end{align}
We conclude, according to  Eq.~\eqref{eq:part_func_HP_path}, that the path free-energy is given by
\begin{eqnarray}
    f_{\text{path}}(\beta) &=& \lim_{N\to\infty} -\frac {1}{N\beta} \log Z_{\text{path}}(\beta)\nonumber  \\ &=& \min_{\mathbf{m}, \mathbf{q}} f_{\text{path}}(\beta,\mathbf{m}, \mathbf{q})\,,
    \label{eq:Zpath}
\end{eqnarray}
where we have defined the free-energy functional
\begin{equation}
    f_{\text{path}}(\beta, \mathbf{m}, \mathbf{q}) = -\frac{1}{2}\sum_{\mu,t}{(m_t^\mu)^2} + \sum_t\pot(q_t) - \frac 1\beta \mathcal{S}(\mathbf{m}, \mathbf{q})\,.
\end{equation}
The minimum of $f_{\text{path}}$ is reached for the roots of
\begin{equation}
    \hat{\mathbf{m}}= \beta\, \mathbf{m}\  ,\ \hat{\mathbf{q}}= -\beta \, \pot'\big(\mathbf{q}\big)\, ,
    \label{eq:seddle_point}
\end{equation}
which, together with Eq.\eqref{eq:der_logZ_1d}, form a closed set of self-consistent equations for the order parameters.

%Since $\hat{\mathbf{m}}$ and $\hat{\mathbf{q}}$ are only approximation of the saddle point in equation~\eqref{eq:entropy}, we need to incorporate the last two derivatives in Eq.~\eqref{eq:entropy_corr} to get a more precise estimation of the entropy.

%\begin{multline}
%   \mathcal{S}(\mathbf{m}, \mathbf{q}) = - \frac{\mathrm{d} f_{\text{path}}}{\mathrm{d} (1/\beta)} = \mathcal{S}({\mathbf{m}},{\mathbf{q}},\hat{\mathbf{m}},\hat{\mathbf{q}}) + \\ \frac{\partial \mathcal{S}}{\partial \hat{\mathbf{m}}}\frac{\partial \hat{\mathbf{m}}}{\partial (1/\beta)} + \frac{\partial\mathcal{S}}{\partial \hat{\mathbf{q}}}\frac{\partial \hat{\mathbf{q}}}{\partial (1/\beta)}\,.
%   \label{eq:entropy_corr}

\subsection{Free-energy for paths}
\label{HP_model}

\begin{figure}
    \centering
    \includegraphics[width=0.7\columnwidth,height=.5\columnwidth]{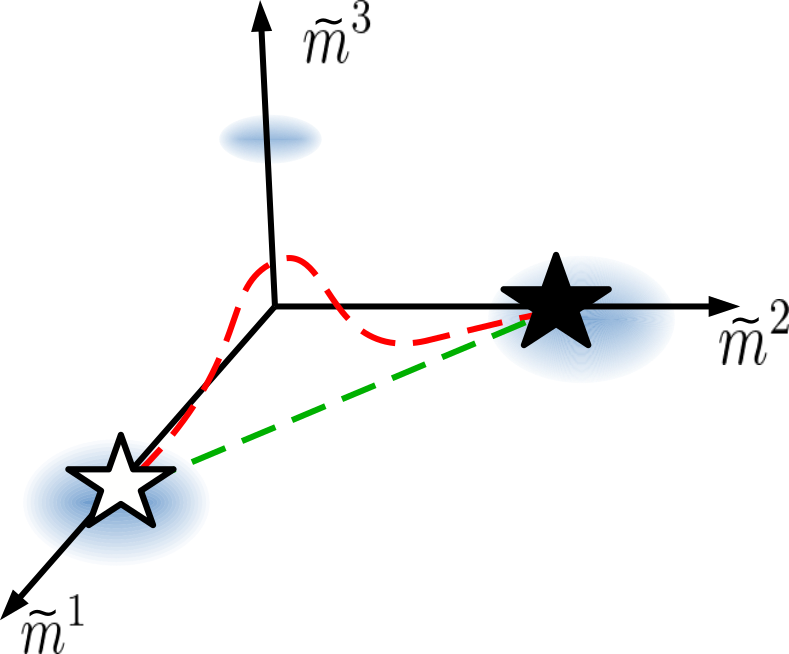}
    \caption{\textbf{Transition paths in the 3D-space of projections $\mathbf{\tilde m}$} for the MHP model presented in Section~\ref{sec:mhp78}. Direct solutions (in green) linearly interpolate in the input space the two minima of the energy landscape. The global solutions are pushed away from the direct ones by the presence of a third minima emerging from the overlap $\omega$ between the two patterns of the model defined in Eq.~\eqref{eq:patterns_HP}.}
    \label{fig:HPsketch}
\end{figure}

To make the theory easier to interpret in the case of the MHP model, we introduce the three projections (denoted by $\tilde{m}_t^\mu$, with $\mu=1,2,3$, along the vectors $\delta_{v_{i},a} $, $\delta_{v_{i},b} $, $\omega \,\delta_{v_{i},c}$, see Fig.~\ref{fig:HPsketch}. While introducing an additional order parameter compared to the number of patterns  makes the computation slightly more  lengthy, it offers the major advantage to allow for immediate distinction between direct ($\tilde m^3=0$) and  global ($\tilde{m}^3 >0$) paths. With this choice, we rewrite the free energy of the path as
\begin{multline}
    f_{\text{path}}(\beta,\tilde{\mathbf{m}},\mathbf{q}) =  \sum_t \bigg( \frac 12 (\tilde{m}^1_t)^2 + \frac 12 (\tilde{m}^2_t)^2 + (\tilde{m}^3_t)^2 + \\+ \tilde{m}^3_t(\tilde{m}^1_t + \tilde{m}^2_t) \bigg) + \sum_t( \pot(q_t) - q_t\pot'(q_t) ) - \frac{1}{\beta} \log {Z}_{\text{1D}}\,,
\end{multline}
where
\begin{multline}
    {Z}_\text{1D} = \sum_{\{v_t\}} \exp\bigg[ \beta\bigg( \sum_t \delta_{v_t,a}(\tilde{m}^1_t + \tilde{m}^3_t) +\delta_{v_t,b}(\tilde{m}^2_t + \tilde{m}^3_t) +  \\ +\omega \delta_{v_t,c}(\tilde{m}^1_t+\tilde{m}^2_t + 2\tilde{m}^3_t)  -\pot'(q_t)\delta_{v_t,v_{t+1}} \bigg)\bigg]     
    \label{eq:partition_global}
\end{multline}

As we shall see, this model undergoes a first order phase transition in the regime where $\beta \times T $ is large controlled by the overlap between patterns, $\omega$, the length of the path, $T$, and the stiffness of the Cont potential, $\gamma$. We will show the existence of a stretched regime when either $T$ and $\omega$ are small or $\gamma$ is large. In this regime the minimum of the free energy corresponds to the direct solution from $\mathbf{v}_{\text{start}}$ to $\mathbf{v}_{\text{end}}$ that one obtains by restricting the sum in ${Z}_\text{1D}$ over the first two colors only. We will refer to this solution as $\#_{\text{dir}}$. If either $T$ and $\omega$ are large or $\gamma$ is small, a floppy regime arises and $\#_{\text{dir}}$ is no longer a minimum of the free energy, and the latter is minimized  by global paths introducing novel mutations at intermediate steps with non zero value of $\tilde{m}^3_t$. 

%at $\omega=\omega_c$ in the limit $\beta \times T \to \infty$. In this limit, when $\omega<\omega_c$, the minimum of the free energy corresponds to the direct solution from $v_0$ to $v_{T}$ that one obtains by restricting the sum in ${Z}_\text{1D}$ over the first two colors only. We will refer to this solution as $\#_2$. When $\omega>\omega_c$, this solution is no longer a minimum of the free energy, and the latter is minimized  by global paths introducing novel mutations at intermediate steps with non zero value of $\tilde{m}^3_t$. 

\subsection{Minimization of the path free-energy in the direct subspace}
\label{SI:HP_minimization}
To understand this phase transition, we first have to find a solution of the direct problem $\#_2$, that is, the set of parameters $\{\tilde{m}^{1,\text{dir}}_t,\tilde{m}^{2,\text{dir}}_t,q^{\text{dir}}_t\}$. The direct solution is found by solving the following coupled equations similar to Eq.\eqref{eq:der_logZ_1d}:
%\begin{widetext}
\begin{align}
    \tilde{m}^{1,\text{dir}}_t&=\frac{1}{{Z}^{\text{dir}}_{\text{1D}}}\sum_{\{v_t=a,b\}} \delta_{v_t,a}\, e^{ -\beta E_{\text{1D}}(\{v_t\})}\ , \label{eq:dir_sol1}\\%\exp\left[ \beta \sum_t \tilde{m}^1_t \delta_{v_t,\mathbf{a}}+\tilde{m}^2_t \delta_{v_t,\mathbf{b}} - \pot'(q_t)\delta_{v_t,v_{t+1}}\right]\\
     \tilde{m}^{2,\text{dir}}_t&=\frac{1}{{Z}^{\text{dir}}_{\text{1D}}}\sum_{\{v_t=a,b\}} \delta_{v_t,b}\, e^{ -\beta E_{\text{1D}}(\{v_t\})}\ , \label{eq:dir_sol1_bis}\\%\exp\left[ \beta \sum_t \tilde{m}^1_t \delta_{v_t,\mathbf{a}}+\tilde{m}^2_t \delta_{v_t,\mathbf{b}} - \pot'(q_t)\delta_{v_t,v_{t+1}}\right]\\
    q^{\text{dir}}_t &= \frac{1}{{Z}^{\text{dir}}_{\text{1D}}}\sum_{\{v_t=a,b\}} \delta_{v_t,v_{t+1}}\, e^{ -\beta E_{\text{1D}}(\{v_t\})}\ . %\exp\left[ \beta \sum_t \tilde{m}^1_t \delta_{v_t,\mathbf{a}}+\tilde{m}^2_t \delta_{v_t,\mathbf{b}} - \pot'(q_t)\delta_{v_t,v_{t+1}}\right]\,, \\
    \label{eq:dir_sol2}
\end{align}
where
\begin{equation}
     E_{\text{1D}} =- \sum_t \bigg(\tilde{m}^{1,\text{dir}}_t \delta_{v_t,a}+\tilde{m}^{2,\text{dir}}_t \delta_{v_t,b} - \pot'(q^{\text{dir}}_t)\delta_{v_t,v_{t+1}}\bigg)\,.
\end{equation}
%\end{widetext}
The partition function ${Z}^{\text{dir}}_{\text{1D}}$ is the same as in~Eq.\eqref{eq:partition_global} with the sum running over the states $a,b$ only, and $\tilde m^3=0$.

We now derive the analytical expression for the mean-field solution when $T\gg 1$ (remember $N$ was sent to infinity first). Due to exchange symmetry $a \leftrightarrow b$ we have $\tilde{m}^{2,\text{dir}}_t = 1 - \tilde{m}^{1,\text{dir}}_t$. We then look for a direct solution of the form
\begin{equation}
    \tilde{m}^{1,\text{dir}}_t= \tilde m \left(\tau = \frac tT\right)\ ,
\end{equation}
where
\begin{align}
& \tilde{m}(\tau) = \begin{cases}
1 &\text{for} \ \tau < \hat{x}\\
1 - \frac{\tau- \hat{x}}{1-2\hat{x}} + \eta(\tau) &\text{for}\ \tau \in (\hat{x},1-\hat{x})\\ 
0 &\text{for}\ \tau > 1-\hat{x}\\
\end{cases}\,
\end{align}
where $\hat x$ depends on $T$ and the function $\eta(\tau)$ vanishes at large $T$; We will show below that $\eta$ is of the order of $1/\sqrt{T}$. 

As the number of mutations at each step $t$ is equivalent to the difference in the projection $\tilde{m}^{1,\text{dir}}$ between steps $t$ and $t+1$, we write
\begin{multline}
    q^{\text{dir}}_{t}= 1 - \frac{\text{nb. mutations}}N = 1 + \tilde{m}^{1,\text{dir}}_{t+1}-\tilde{m}^{1,\text{dir}}_{t} \\ =  1 + \frac 1T\,\partial_\tau \tilde{m}(\tau)\, 
\end{multline}
to dominant order in $T$. Hence the overlap order parameters are fully determined once the projection is, with the explicit expression $q^{\text{dir}}_{t}=q(\tau = t / T)$ and
\begin{align}
 &q(\tau ) = \begin{cases}
1 &\text{for} \ \tau < \hat{x}\\
1 + \frac{1}{T}\left(\frac{-1}{1-2\hat{x}} +  \eta'(\tau)\right) &\text{for}\ \tau\in (\hat{x},1-\hat{x})\\ 
1 &\text{for}\ \tau> 1-\hat{x}\ .\\
\end{cases}\,,
\end{align}

Our goal is to inject the above Ans\"atze into Eq.~\eqref{eq:dir_sol2} and determine the function $\eta$ and the value of $\hat{x}$ that solve the equation at the $0$-th order in $T$. First, we expect the effective coupling $-\pot'(q(\tau))$ between neighbouring $v_t,v_{t+1}$ in the energy $E_{\text{1D}}$ to scale linearly with the size of the system $T$. The reason is that, given a configuration $\{v_t\}$ appearing in the sum of ${Z}^{\text{dir}}_{\text{1D}}$, every couple of adjacent sites $v_t$ and $v_{t+1}$ occupying different states, {\em i.e.} for every mutation along the path would produce an energetic penalty $- \pot'(q_t)\delta_{v_t,v_{t+1}}$ of the order of $T$. The partition function will thus be dominated by the configurations $v_t=a$ for $\tau<\hat{x}$ and $v_t=b$ for $\tau >1-\hat{x}$, that is, by configurations with a single mutation along the path. 

Computing the derivative of the Cont potential, we obtain $-\pot'(q(\tau)) = |\gamma-1/(1-2\hat{x})+\eta'(\tau)|^{-2}$. Therefore, we expect 
\begin{equation}\label{eq:eta1}
\gamma-\frac 1{1-2\hat{x}}+\eta'(\tau)\equiv  \frac{\xi(\tau)}{\sqrt T} \ . 
\end{equation}
The partition function can then be rewritten as 
\begin{multline}
   {Z}^{\text{dir}}_{\text{1D}} = T \int_0^1 \mathrm{d}\tau \exp \Bigg[\beta T\Bigg( \int_0^\tau\mathrm{d}y\, \tilde{m}(y) +\\+\int_\tau^1\mathrm{d}y\, (1-\tilde{m}(y)) - \frac{1}{\xi(\tau)^{2} } \Bigg) \Bigg]\, 
   \label{eq:part_dir}
\end{multline}
where we explicitly integrate over the reduced `time' $\tau$ at which the $a\rightarrow b$  mutation occurs. When $\beta T\gg 1$, the exponential integral in the partition function should not depend on $\tau$ as the mutation may take place with uniform probability in the interval $(\hat{x}, 1-\hat{x})$; hence, the mutations will happen at different times depending on the site $i$. Differentiating the term in factor of $\beta T$ with respect to $\tau$ we obtain the following differential equation for $\tau \in (\hat{x}, 1-\hat{x})$:
\begin{equation}
    \tilde m(\tau) - \big( 1- \tilde m (\tau)\big)  -\frac {d}{d\tau} \left(\frac{1}{\xi(\tau)^{2} } \right) =0\, ,
\end{equation}
or, equivalently in the large $T$ limit,
\begin{equation}
    1- 2 \frac{\tau - \hat{x}}{1-2\hat{x}}   + 2\frac{\xi'(\tau) }{\xi(\tau)^{3} } =0\,.
\end{equation}
Solving this differential equation leads to
\begin{equation}
    \xi(\tau) = \left[ \frac{1}{\xi(\hat{x})^{2}} - \frac{\tau^2-\hat{x}^2 - (\tau-\hat{x})}{(1-2\hat{x})} \right]^{-\frac{1}{2}}\,.
\end{equation}
In order to ensure the continuity of $\pot'(q(\tau))$ in $\tau=\hat{x}$, we choose $\xi(\hat{x})=\gamma \sqrt{T}$. Integrating Eq.\eqref{eq:eta1} over $\tau$ we obtain
\begin{equation}
    \eta(\tau)-\eta(\hat{x}) = \frac{1}{T^{1/2}} \int_{\hat{x}}^\tau \xi(y) \mathrm{d} y +\left(\frac{1}{1-2\hat{x}} - \gamma\right)(\tau-\hat{x}) \,.
    \label{eq:self_cons}
\end{equation}
Last of all, upon imposing the boundary condition $\eta(\hat{x}) = \eta(1-\hat{x}) = 0 $, we also determine $\hat{x}$ as a function of $\gamma$ and of $T$. In particular, we can expand $\hat{x}$ for large $T$ as
\begin{equation}
    \hat{x} = \frac 12 - \frac{1}{2\gamma} -\frac{\pi}{2\sqrt{\gamma^3 T}}  + o\big(T^{-\frac 12}\big)\, .
    \label{eq:scale_xhat}
\end{equation}
Consequently, $\tilde{m}(\tau)=\tilde{m}^{\infty}(\tau) +\mathcal{O}\left(T^{-\frac 12}\right)$ with $\tilde m^{\infty}(\tau)=1$ if $\tau<\hat x^{\infty}=\frac 12 \big(1-\frac 1\gamma\big)$, $\tilde m^{\infty}(\tau)=0$ if $\tau>1-\hat x^{\infty}$, and
\begin{equation}\label{eq:mtt2}
     \tilde{m}^{\infty}(\tau)=1-\gamma\left(\tau -\hat x^{\infty}\right) \ \, 
\end{equation}
if $\hat x^{\infty} \le \tau \le 1-\hat x^{\infty}$. It is easy to check that Eqs.\eqref{eq:dir_sol1},\eqref{eq:dir_sol2} are fulfilled at zeroth order by this solution. 

\begin{figure}
    \centering
    \includegraphics[width=0.8\linewidth]{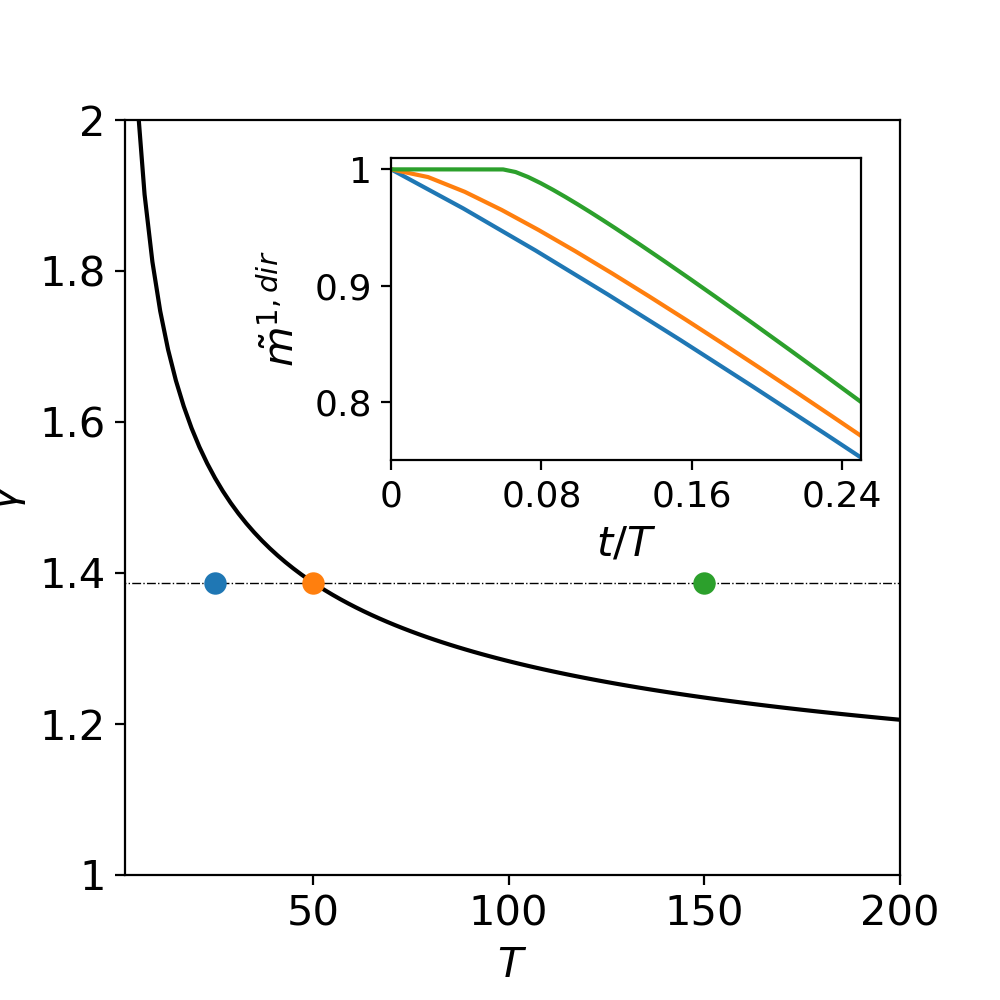}
    \caption{{\bf The `understretched' and `overstretched' sub-regimes for direct paths}. The solid black line represents the root $\gamma^*(T)$  of Eq.~\eqref{eq:strec_over_trans}. The three colored dots on the black dashed line $\gamma = 1.387$ correspond to $T=25$ (blue), $50$ (orange), $150$ (green). The blue and green dots respectively correspond to the overstretched ($\hat{x}=0$, $\gamma < \gamma^*(T)$) and understretched ($\hat{x}>0$, $\gamma > \gamma^*(T)$) direct regimes. The orange dot locates the crossover point ($\gamma = \gamma^*(T)$).  Inset: Numerical solutions for $\tilde{m}^{1,\text{dir}}_t$ with those combinations of parameters are shown in the inset plot for $t/T\le 0.24$. In the simulations $\beta=6$.}
    \label{fig:gammastar}
\end{figure}

\begin{figure*}
    \centering
    \includegraphics[width=.9\textwidth]{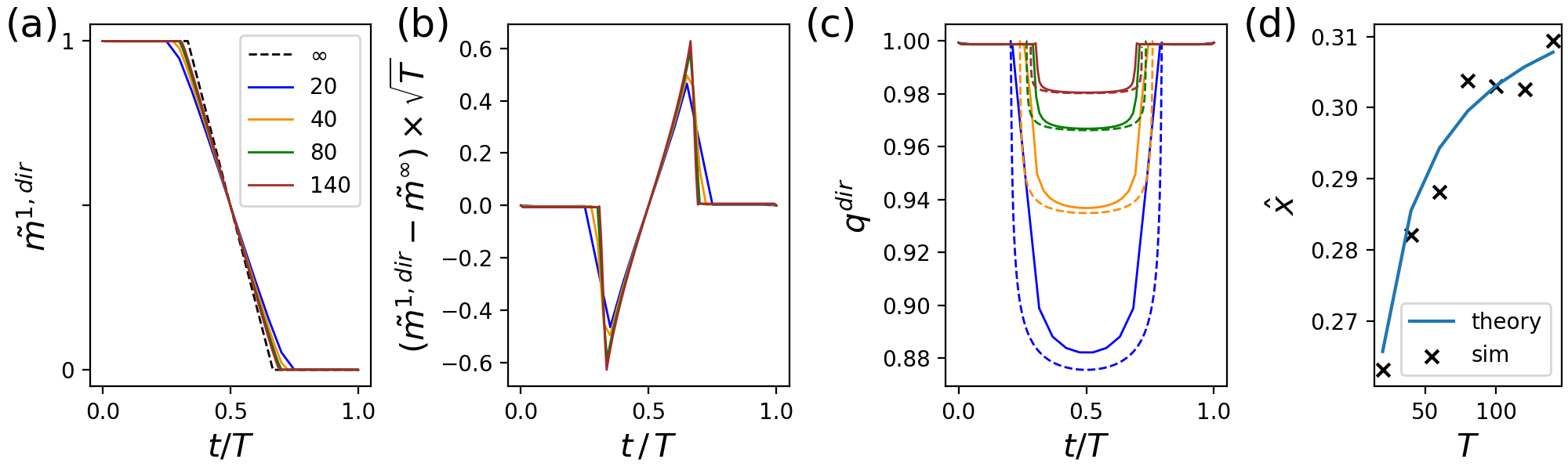}
    \caption{\textbf{Mean-field solution of the MHP model in the understretched regime for direct paths}. (a) Numerical solutions for $\tilde{m}^{1,\text{dir}}_t$ for different values of $T$ (values showed in legend) compared with the limit solution $\tilde{m}^{\infty}$ in Eq.(\ref{eq:mtt2}) for $T\to\infty$ (black dashed line). (b) Scaling of the difference between $\tilde{m}^{1,\text{dir}}_t$ computed numerically and $\tilde{m}^{\infty}_t$ for large $T$. (c) Numerical solutions for $q^{\text{dir}}_t$ (solid lines) compared with the respective theoretical estimation (dashed lines) evaluated using $\hat{x}$ according to Eq.\eqref{eq:scale_xhat}. (d) Numerical estimation of $\hat{x}$ (black crosses: the value corresponds to the moment $\tilde{m}^{1,\text{dir}}_t$ becomes $<1$) vs. theoretical scaling from Eq.\eqref{eq:scale_xhat} (blue line). The parameters of the simulations are $\beta=6$, $\gamma=3$.}
    \label{fig:HP_model_sol}
\end{figure*}

The solution above holds as long as $\hat{x}$ does not hit the boundary, {\em i.e.} provided $\hat{x} > 0$. When $\hat x=0$, using Eq.\eqref{eq:self_cons} and integrating function $\xi$, we find that $\gamma$ has to satisfy the equation 
\begin{equation}
    \gamma = 1 + \frac 2{\sqrt{T}} \arctan{\left(\frac{\gamma\sqrt{T}}{2}\right)}\ .
    \label{eq:strec_over_trans}
\end{equation} 
The root of this equation, which we denote by  $\gamma^*(T)$ is plotted in Figure~\ref{fig:gammastar}. We may now conclude:
\begin{itemize}
    \item If $\gamma < \gamma^*(T)$ we have $\hat{x} = 0$: the projection $\tilde m (\tau)$ is smaller than 1 as soon as $\tau>0$, see inset in Figure~\ref{fig:gammastar}. For such small $\gamma$ the paths are not flexible enough and the full `time' $T$ at their disposal is needed to join the anchoring edges. We call this regime {\em overstretched}. Notice that the boundary conditions $\eta(\hat{x}=0)= 0$ in Eq.~\eqref{eq:self_cons} can be satisfied by fixing the initial value of the function $\xi$, \emph{i.e.} $\xi(0)$. In particular, we find
\begin{equation}\label{eq:xi0}
    \xi(\hat{x}=0) = 2 \tan \left(\frac{\sqrt{T} (\gamma-1)}{2}\right)\,.
\end{equation}
\item If $\gamma > \gamma^*(T)$, we have $\hat x >0 $. The available number of intermediate sequences along the path, $T$, is larger than what is actually needed to join the two edges. A fraction ($=2\hat x$) of these intermediate sequences are mere copies of  the initial and final configurations, see inset of Figure~\ref{fig:gammastar}. We hereafter call this regime {\em understretched}. All the analytical results reported in Eqs.~(\ref{eq:scale_xhat},\ref{eq:mtt2}) are in excellent agreement with the numerical resolution of the self-consistent equations for the order parameters, see Figure~\ref{fig:HP_model_sol}.
\end{itemize}

\subsection{The direct-to-global phase transition}
\label{SI:HP_direct_to_global}

The solution $\#_{\text{dir}}$ we have derived above assumes that $\tilde m_3$ vanishes at all time. This assumption is correct as long as the minimum of the free energy $f_{\text{path}}$ is located in $\tilde m_3=0$.
We compute below the first derivative of the free energy along the third projection $\tilde{m}^3_t$:
\begin{equation}
    \left. \frac{\partial f_{\text{path}}}{\partial \tilde{m}^3_t } \right|_{\#_{\text{dir}}} = 1 - \left. \langle \delta_{v_t,a} + \delta_{v_t,b}+2\omega\delta_{v_t,c} \rangle_\text{1D}\right|_{\#_{\text{dir}}}\,.
    \label{eq:derM}
\end{equation}
By studying the sign of this derivative we will show the existence of a critical value of $\omega$ appearing in the patterns of the HP model, see Eq.~(\ref{eq:patterns_HP}). This critical value, hereafter denoted by $\omega_c$, separating a regime where the direct solution is stable ($\omega<\omega_c$) and a regime where it is not and the true mean-field solution is global ($\omega>\omega_c$).

Two classes of competing configurations must be considered: the direct ($dir$) ones, which start in $v^{\text{start}}=a$ and turn into $v^{\text{end}}=b$ at some time $\tau\in(\hat{x},1-\hat{x})$.; the global ($glob$) ones, which start in $a$ then change to $c$ at some time $\tau\equiv x \in(0,1/2)$, then turn into $b$ when $\tau=1-x$. We estimate below the energies $E_{\text{dir}}$ and $E_{\text{glob}}$ corresponding to the two scenarios. In particular, when $E_{\text{dir}}<E_{\text{glob}}$, the direct configurations dominate the average on the right hand side of Eq.~\eqref{eq:derM}, leading to 
\begin{equation}
\left.\frac{\partial f_{\text{path}}}{\partial \tilde{m}^3_t } \right|_{\#_{\text{dir}}} = 0\ \forall\ t\,.
\end{equation}
Conversely, when $E_{\text{dir}}>E_{\text{glob}}$, we will have
\begin{equation}
    \left.\frac{\partial f_{\text{path}}}{\partial \tilde{m}^3_t } \right|_{\#_{\text{dir}}} = 1 - 2\omega\ \text{for}\ t\in(x,1-x)\, ,\ 0\ \text{otherwise}.
\end{equation}
Hence, the direct solution will be unstable if, in addition, $\omega>\frac 12$. As we shall check explicitly below this condition is always met when $E_{\text{dir}}>E_{\text{glob}}$.

\subsubsection{Understretched regime}

The energy of the direct configurations (for $T\gg 1$) is given by:
\begin{equation}
    E_{\text{dir}} = -T \left( \hat{x} + \frac{1}{2}  \right) + \frac{1}{\gamma^{2}}\,,
\end{equation}
while the global ones have energy
\begin{equation}
    E_{\text{glob}}(x)= \left\{
    \begin{aligned}
    &-T \left( 2x + \omega(1-2x)  \right) + \frac{2}{\gamma^{2}} \quad \text{for}\ x\le\hat{x}\\
    &-T \Big( 2\hat{x} + 2\int_{\hat{x}}^x\mathrm{d}y\,(1-\frac{y-\hat{x}}{1-2\hat{x}}) + \\&+\omega(1-2x) - \frac{2}{|\xi(x)|^{2}} \Big) \quad \text{for}\ x\in(\hat{x},1/2)
    \end{aligned}
    \right.\,
\end{equation}
which is minimal for $x=\hat{x}$ when $\omega\in(1/4,1)$ and for $x=0$ when $\omega>1$. Here the condition $E_{\text{glob}}<E_{\text{dir}}$ provides the critical value of $\omega$ for the phase transition: 
\begin{equation}
    \omega^{\text{under}}_c (\gamma,T) = \frac 12 + \frac 1{T\gamma^{2}(1-2\hat{x})} \simeq \frac 12 + \frac 1{T\gamma}\  
\end{equation}
for large $T$.

\subsubsection{Overstretched regime}

In the overstretched case, the energy of the direct configurations is given by
\begin{equation}
    E_{\text{dir}} = - \frac{T}{2} + \frac{T}{\xi(0)^{2}}\ ,
\end{equation}
while the global configurations correspond to  energy 
\begin{equation}
    E_{\text{glob}} = - T \omega  + \frac{2T}{\xi(0)^{2}}\,.
\end{equation}
Here, $\xi(0)$ is given by Eq.\eqref{eq:xi0}.
The condition $E_{\text{glob}} < E_{\text{dir}}$ leads to a new critical value for $\omega$:
\begin{equation}
  \omega^{\text{over}}_c (\gamma,T) = \frac 12 + \frac 1{4 \tan^2(\sqrt{T}(\gamma-1)/2)}\ .
\end{equation}

\subsubsection{Comparison with numerics}

\begin{figure}
    \centering
    \includegraphics[width=\linewidth]{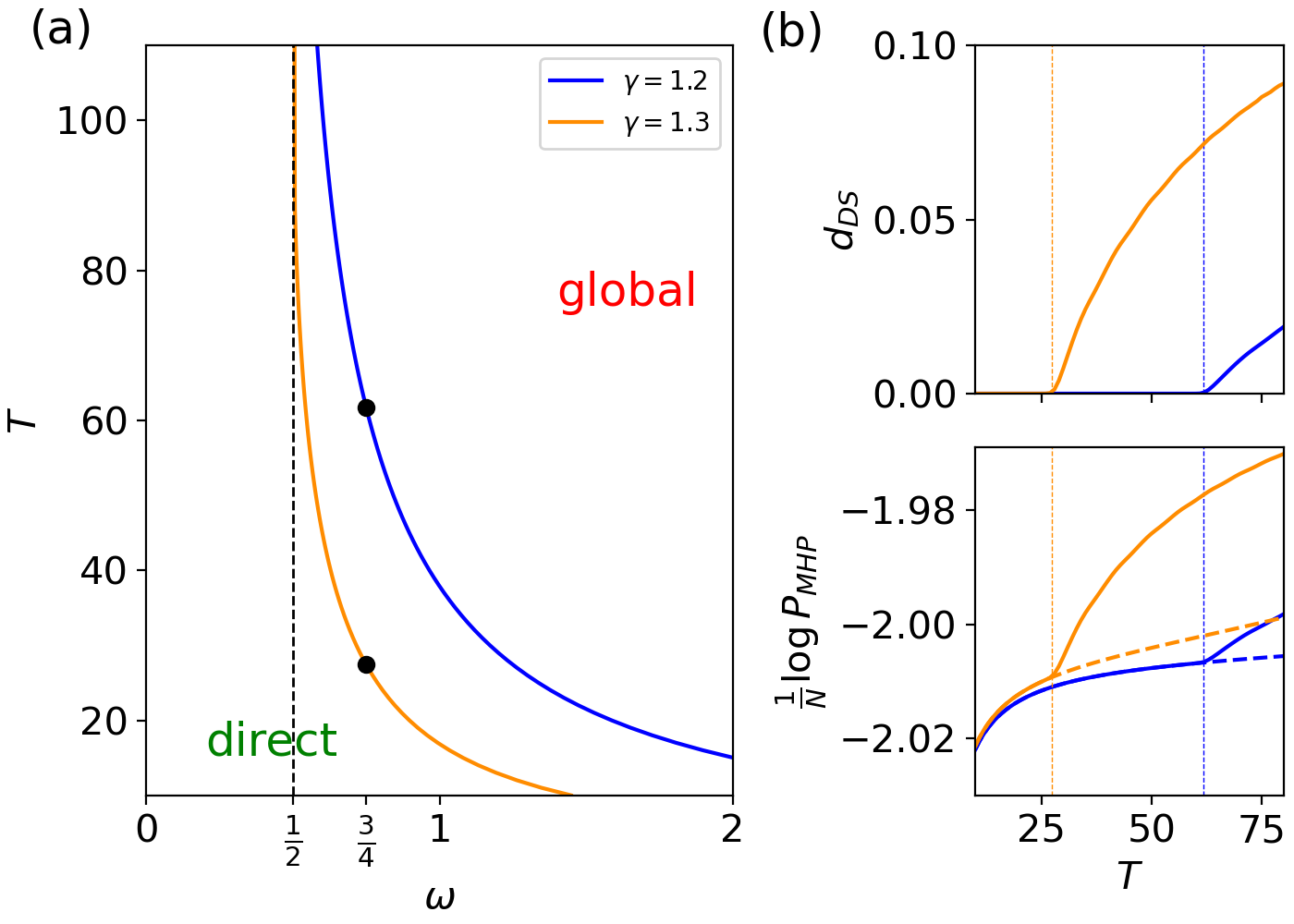}
    \caption{{\bf Crossover between direct and global transition paths in the MHP model.} (a). Critical line $\omega_c(\gamma,T)$ vs. $T$ for two values of  $\gamma$, see Eq.~\eqref{eq:tco}. The black dots show the crossovers for $\omega=\frac 34$.  {(b)} Distance $d_{\text{DS}}$ to the direct space (Top) and $(\log P_{\text{MHP}})/N$ averaged over intermediate sequences  (Bottom; solid line: global, dashed: direct)  vs. path length $T$; same parameters as in (a). }%(c)  Mean-field estimates of $d_{\text{DS}}$ (Top) and of $(\log P_{\text{RBM}})/N$ (Bottom; red: global paths, green: direct) vs. $\gamma$ for mutational paths of the WW domain of length $T=10$. In all panels $\beta=3$. }
    \label{fig:MFlike}
\end{figure}

Putting together the two regimes studied above, we find that the transition takes place at
\begin{equation}
    \omega_c (\gamma,T)= \left\{  \begin{aligned}
        &\omega_c^{\text{over}}(\gamma,T)\ &\text{for}\ \gamma < \gamma^*(T)\\
        &\omega_c^{\text{under}}(\gamma,T)\ &\text{for}\ \gamma > \gamma^*(T)\\
    \end{aligned}\right.\,.%\max \left(  \omega_c^{\text{over}}(\gamma,T) ,\omega_c^{\text{under}}(\gamma,T)  \right)\, .
    \label{eq:tco}
\end{equation}
The phase diagram in the $(\omega,T)$ plane is shown in Figure~\ref{fig:MFlike} for different values of the flexibility parameter $\gamma$.

\begin{figure}
    \centering
    \includegraphics[width=\linewidth]{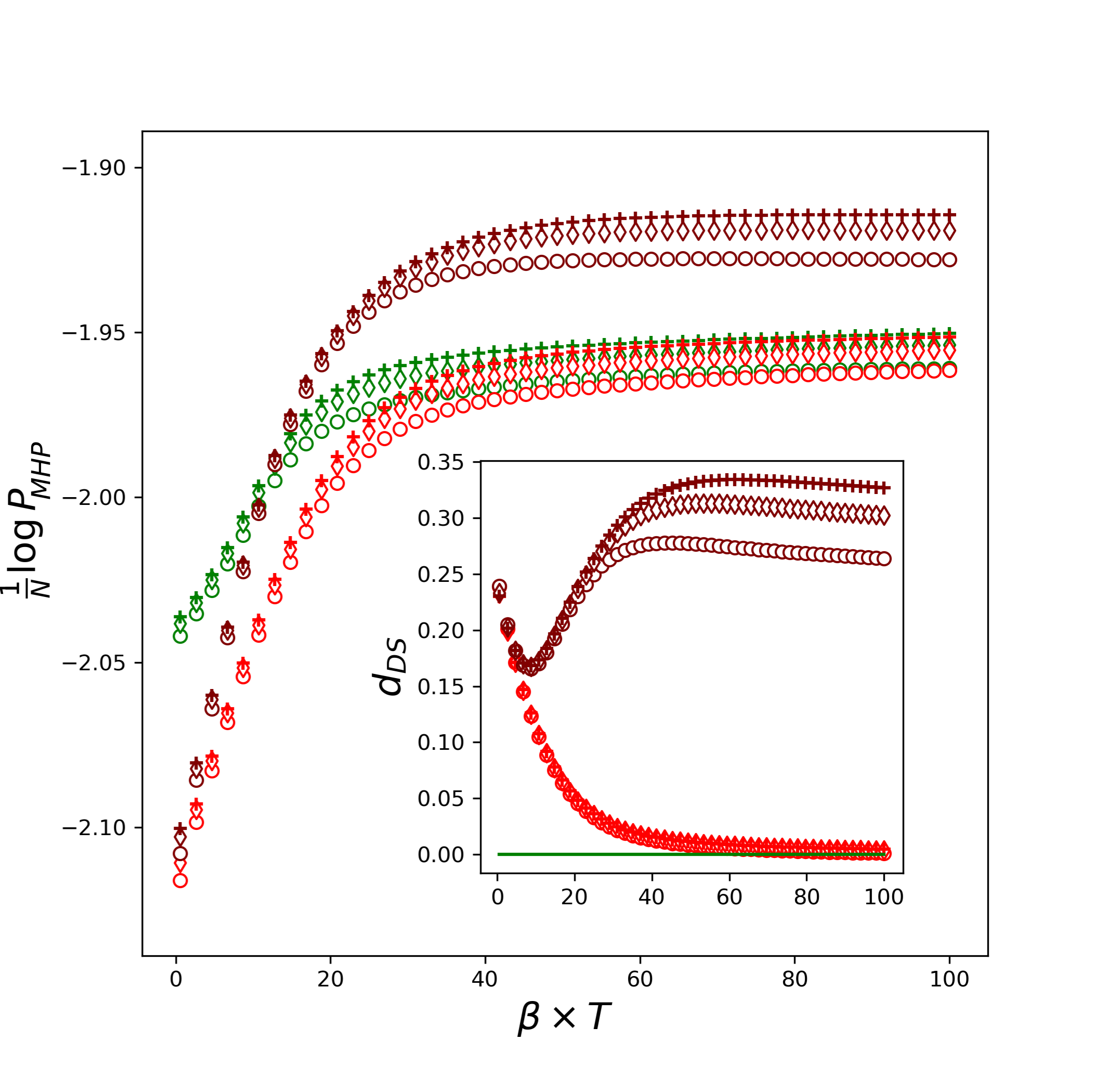}
    \caption{\textbf{Average log-likelihood along the paths for the MHP model} as a function of $\beta \times T$. Inset plot shows the average distance to direct space. Symbols stands for different $T$ (circles for $T=20$, diamonds for $T=30$ and pluses for $T=40$). Green symbols represents direct solutions (which are of course independent of $\omega$), Red symbols represents global solutions with $\omega=0.4$ and maroon symbols represent global solutions for $\omega=0.7$. Here  $\gamma=2$ (Cont potential). For high values of $\beta \times T$ we see that the global paths for $\omega<0.5$ converge towards the direct ones, while, for $\omega > 0.5$, the two classes of paths remain separated, in agreement with the phase transition shown in Fig.~\ref{fig:MFlike}.}
    \label{fig:hop_beta}
\end{figure}

While the transition formally takes place in the limit $\beta\times T\to \infty$, a cross-over is observed for finite $T$ and $\beta$. We show in  Figure ~\ref{fig:hop_beta} the coincidence of the average log-likelihoods of intermediate sequences along direct and global paths at large $T$ for small $\omega$, and the higher quality of global paths for large $\omega$. Notice that these results are valid when $T$ is sent to large values while keeping $\beta$ fixed. If $\beta$ is small, e.g. of the order of $\frac 1T$, the domination of global paths on direct paths is due to the larger entropy of the former. Figure \ref{fig:hop_beta} shows that, for small $\beta\times T$, global paths are indeed of lesser quality (probability) than their direct counterparts, even at high $\omega$. 

To better distinguish global from direct paths, we introduce the distance
\begin{eqnarray}
    d_{\text{DS}}(\mathbf{v}) &=& \frac{1}{N} \sum_i (1-\delta_{(v_{\text{start}})_i,v_i})(1-\delta_{(v_{\text{end}})_i,v_i})\ . 
%    \\
%    &=& 1+ q\big(\mathbf{v}^{\text{start}}, \mathbf{v}^{\text{end}}\big)-q\big(\mathbf{v}, \mathbf{v}^{\text{start}}\big)-q\big(\mathbf{v}, \mathbf{v}^{\text{end}}\big)\nonumber 
\end{eqnarray}
By definition, $d_{\text{DS}}$ vanishes if the configuration is within the direct subspace, and is strictly positive otherwise. Its maximal value is $1$.
We show in Fig.~\ref{fig:hop_beta}(inset) the behavior of $d_{DS}$ for two values of the flexibility parameter controlling the Cont potential, below and above the transition point.

\subsection{Escaping from local minimum: paths anchored at origin}
\label{sec:escape_HP}

We have so far considered paths anchored at both extremities. Our mean-field formalism can be extended to the case of paths in which the final configuration is not fixed. In this context the goal is to  characterize the most likely behavior of a path in the energy landscape under a mutational dynamics encoded in the interaction potential $\Phi$. 

A natural question in this scenario is to estimate when and in which conditions a configuration escape from a local minimum to reach a more stable configuration. In the MHP model, we consider paths starting in ${\bf v}_{\text{start}} = \{ a,a,a, \dots, a\}$, and unconstrained at the other extremity. The properties of these paths can be computed through Eq.~\ref{sec:MF}, upon relaxing the condition at the extremity when computing $Z_i^{\text{1D}}$ using transfer matrix in Eq.~\eqref{eq:logZ_1d}.

\begin{figure}
    \centering
    \includegraphics[width=0.9\linewidth]{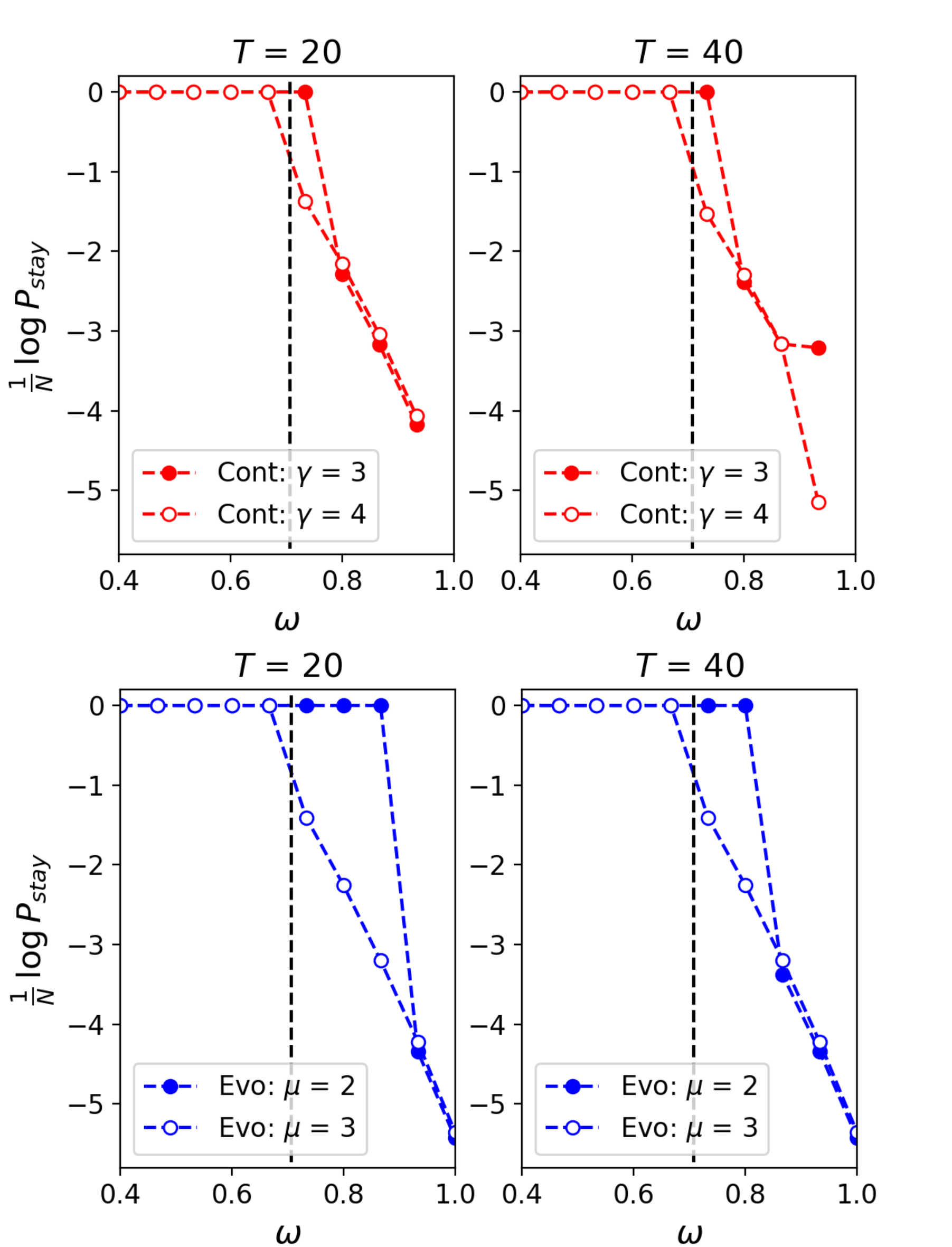}
    \caption{{\bf Probability of stay in the initial local minimum} close to the configuration ${\bf m}^{\text{start}}=(1,0)$ after $T=20$ (\emph{left}) and $T=40$ (\emph{right}) steps. the probabilities are plotted against different values of the overlap $\omega$. Cont (\emph{up}) and Evo (\emph{bottom}) scenario are respectively plotted in red and blue. The dark dashed line correspond to $\omega =1 / \sqrt{2}$, when the minimum at ${\bf m}=(\omega,\omega)$ becomes global. Here $\beta=6$. }
    \label{fig:escape_HP}
\end{figure}

To estimate the escape probability, we define the region $\mathcal{R}$ associated to the minimum of the free-energy landscape close to the initial configuration ${\bf m}^{\text{start}}=(1,0)$. Then, we evaluate the probability $P_\text{stay}$ of remaining in that region after a certain number of steps $T$ using Eq.~\eqref{eq:escape_prob_HP}. The escape probability is computed as $P_{\text{escape}} = 1 - P_{\text{stay}}$. In Figure~\ref{fig:escape_HP} we show the estimated $\log P_\text{stay}$ in the Cont and Evo scenarios for different values of $\omega$ and $T$. When $\omega>1/\sqrt{2}$ the local minimum in $(\omega,\omega)$ depicted in Fig.~\ref{fig:sketch_HP_over} becomes global, and the path is attracted towards this minimum. For finite $T$ higher values of $\omega$ are required to overcome the elastic constraint due to $\Phi$ to remain close to ${\bf v}_{\text{start}}$. 

\section{Paths in data-driven protein models}
\label{sec:inferred}

In this section, we aim to expand our mean-field analysis 
to model the landscapes inferred by the Restricted Boltzmann Machine from data.
Data configuration ${\bf v}$ are sequences  of the same protein family given in a multi-sequence alignment (MSA) of length $N$ and an alphabet of size $A = 21$ (20 amino acids plus the gap symbol). The sequences with high probabilities according to the inferred model are predicted to have high fitnesses.

\subsection{Mean-field theory}
\label{sec:mfrbm}

Due to the bipartite structure of their interaction graph, the mean-field theory of the Hopfield-Potts model presented in Section~\ref{sec:MF} can be easily extended to the case of RBM. Two differences are: (1) the effective energy is not a quadratic function of the projections $m^\mu_t$ when the hidden potentials ${\cal U}(h)$ are not quadratic in $h$; (2) the 1D partition function now depends on the potentials $g_i(v_i)$ acting on the visible units. The expression for the path free energy is now
\begin{equation}
f_{\text{path}}(\mathbf{m},\mathbf{q}) = - \frac 1N \sum_{\mu,t} \Gamma_\mu(N m^\mu_t) +\sum_t \pot(q_t)\, -\frac 1\beta
{\cal S} (\mathbf{m},\mathbf{q}) \ ,
\label{eq:free_en_RBM}
\end{equation}
%\\-&\frac{1}{\beta}  \min_{\{\hat{m}^\mu_t\},\{\hat{q}_t\}}\left[ - \sum_{\mu,t} \hat{m}^\mu_t {m}^\mu_t -\sum_t \hat{q}_t {q}_t + \frac 1N \sum_i \log Z_i\right]\,. \nonumber
where $\Gamma_\mu$ is defined after Eq.~\eqref{eq:tyujk} and the entropy ${\cal S}$ is given by Eq.~\eqref{eq:entropy} with
\begin{multline}
    Z_i^{\text{1D}}=  \sum_{\{v_t\}}\exp \left(\beta\sum_t g_i(v_{t}) \, +\right. \\ \left.+ \sum_{t,\mu} \hat{m}^\mu_t\, {w_{i\mu}(v_t)} +\sum_t\hat{q}_t \,\delta_{v_t,v_{t+1}} \right)\ . 
\end{multline}
$Z^\text{1D}_i$ can be efficiently estimated through products of $A\times A$-dimensional transfer matrices, where $A$ is the number of
Potts states. For global paths, $A = 21$, while $A = 2$ for direct paths. This mean-field theory is exact when $N\to \infty$ \footnote{Note that in order for $\Gamma_\mu$ to be well defined in this limit, we are formally supposing that the local fields on the hidden space are scaling as $N$, \emph{i.e.} $\gamma_{\mu,\pm},\ \theta_{\mu,\pm} = \mathcal{O}(N)$. The weights do not scale with $N$, \emph{i.e.} $w_{i\mu},\ g_i=\mathcal{O}(1)$. } and the numbers of hidden units, $M$, and of steps, $T$ remain finite, but it is already an accurate approximation for some finite-$N$ cases, as will be shown below. 

Once the mean-field solution has been determined through minimization of $f_{\text{path}}$ we can compute any observable, such as the average frequencies of amino acids on site $i$ at intermediate step $t$ on the path:
\begin{align}
    \langle \delta_{v_{i,t},a} \rangle &= \frac{\partial f_{\text{path}}}{\partial (\beta g_{i,t}(a))} = \sum_{\{v_{t'}\}}\frac{\delta_{v_{i,t},a}}{Z_i}  \exp \left(\beta\sum_{t'} g_{i,t'}(v_{t'}) \right. \nonumber\\
    & + \left. \sum_{t',\mu} \hat{m}^\mu_{t'}\, {w_{i\mu}(v_{t'})} +\sum_{t'}\hat{q}_{t'} \,\delta_{v_{t'},v_{t'+1}} \right)\ ,
    \label{eq:freq}
\end{align}
where $\hat{m}^\mu_{t} = \beta \Gamma'_\mu(N{m}^\mu_{t})$ and $\hat{q}_{t} = - \beta \pot'(q_t)$, see Eq.\eqref{eq:seddle_point}.

\subsection{Application to sequence data from Lattice Protein models}
\label{sec:prot1w}

%We now apply our mean-field theory to characterise global and direct paths.
We start by considering the toy-model of Lattice Proteins (LP)~\cite{Lau1989Oct}. The model considers sequences of $N=27$ amino acids that may fold in one out of $\sim 10^5$ possible 3-dimensional conformations, defined by all possible self-avoiding walks going through the nodes of the $3\times 3 \times 3$ cubic lattice. 

Given a structural conformation $\mathbb{S}$, the probability of a sequence $\vsec$ to fold into that structure is given by the interaction energies between amino acids
in contact in the structure (occupying neighbouring nodes on the lattice). In particular, the total energy of sequence $\mathbf{v}$ with respect to structure $\mathbb{S}$ is given by
\begin{equation}
    \mathcal{E}_{\text{LP}}(\mathbf{v}|\mathbb{S}) = \sum_{i<j} c_{ij}^\mathbb{S}E_{MJ}(v_i,v_j)\, ,
\end{equation}
where $c^\mathbb{S}$ is the contact map ($c_{ij}^\mathbb{S}=1$ if sites are in contact and $0$ otherwise), while the pairwise energy $E_{MJ} ( v_i , v_j )$ represents the amino-acid physico-chemical interactions given by the the Miyazawa-Jernigan knowledge-based potential~\cite{Miyazawa1996Mar}. The probability to fold into a specific structure is written as
\begin{equation}
    p_{\text{nat}}(\mathbb{S}|\mathbf{v}) = \frac{e^{-\mathcal{E}_{\text{LP}}(\mathbf{v}|\mathbb{S})}}{\sum_{\mathbb{S}'}e^{-\mathcal{E}_{\text{LP}}(\mathbf{v}|\mathbb{S}')}}\, ,
\end{equation}
where the sum tuns over the entire set of folds on the cubic lattice. The function $p_{\text{nat}}$ represents a suitable landscape that maps each sequence to a score measuring the quality of its folding. 

\begin{figure}
\centering
\includegraphics[width=\linewidth]{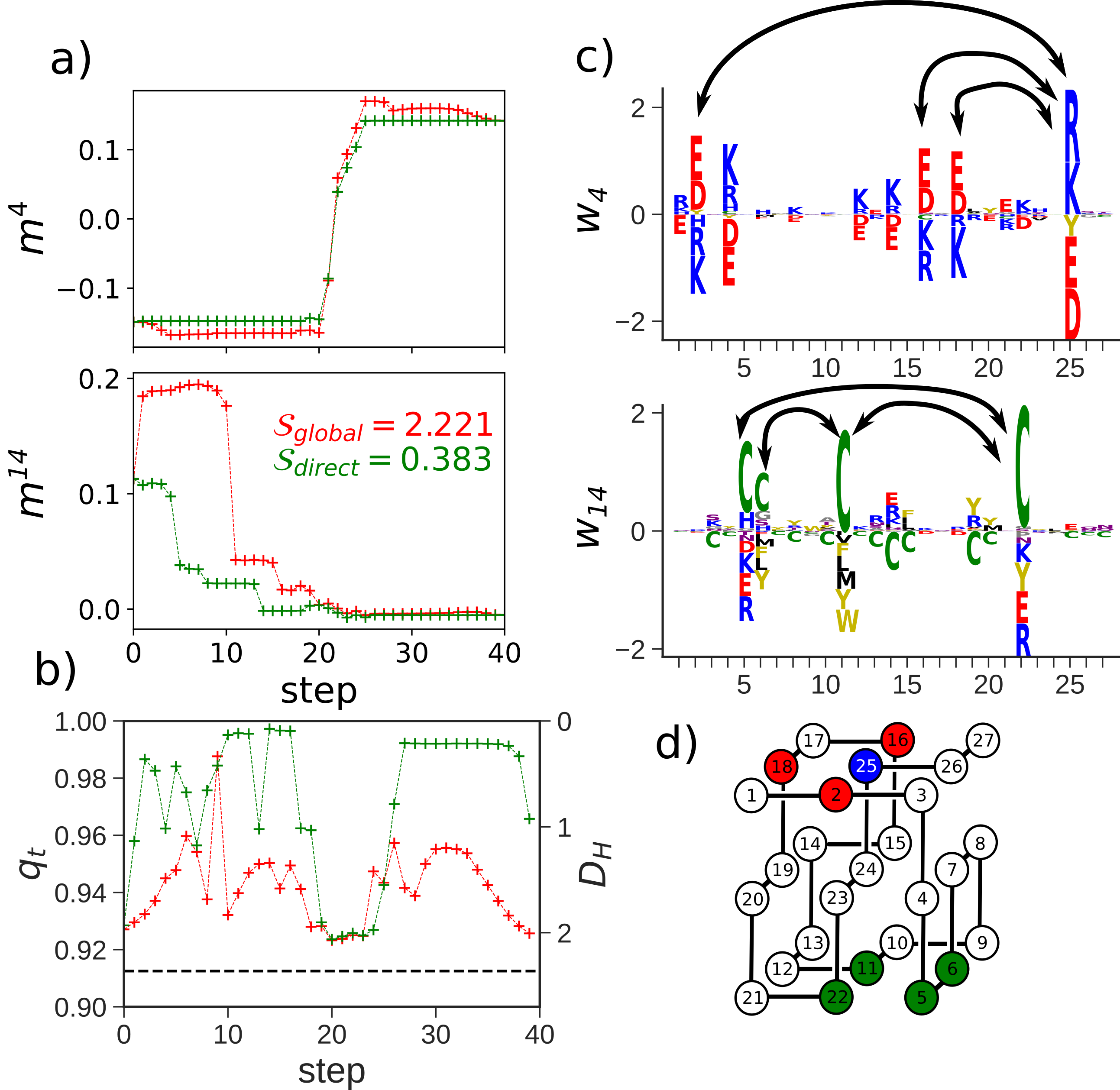}
\caption{{\bf Mean-field description of mutational paths} in lattice proteins with the Cont potential.
{(a)} Values of two relevant inputs vs. number $t$ of mutations along paths of length $T=40$. Red and green lines correspond to, respectively,  global and direct paths.  Parameters: $\beta = 3$,  $\gamma=3.5$. In the inset we show the entropy for the global and direct solutions. (b) Overlap $q_t$ (left scale) and average number of mutations $D_H=N(1-q_t)$ (right scale) between sequences at steps $t$ and $t+1$ vs. $t$; The dark line shows $q_c$. {(c)} Logos of the attached weights $w_{i,\mu}(v)$. Positively charged amino acids are in blue, negatively charged ones in red. {(d)} Reference structure for the Lattice Protein model.}
\label{fig:meanfield}
\end{figure}

To test our mean-field theory, we first train a RBM over sequences sampled from the probability distribution $\propto p^{\beta_s}_{\text{nat}}(\cdot|\mathbb{S})$ for a specific structure $\mathbb{S}$ (with $\beta_s=10^3$) using Monte-Carlo~\cite{Jacquin_LP_2016}. Then we numerically compute the MF solutions for paths connecting two far away target sequence with high $p_{\text{nat}}$ for both the global and direct cases:
\begin{itemize}
    \item[]$\vsec_{\text{start}}\,=\,$\texttt{DRGIQCLAQMFEKEMRKKRRKCYLECD} \ ,
    \item[]$\vsec_{\text{end}}\,=\,$\texttt{RECCAVCHQRFKDKIDEDYEDAWLKCN}.
\end{itemize}
These two configurations are characterized  by a flip of the  charge (from negative to positive) of the amino-acids in the site 25 ( from E to K {}) and of the neighboring sites (see Fig.~\ref{fig:meanfield} d) to keep an attractive interaction between such sites in order to guarantee the stability of the fold.
The trajectories of the inputs $m^\mu _t$ and of the overlaps $q_t$ reveal which and when latent factors of RBM enter into play in the transition. 

Figure~\ref{fig:meanfield}(a) shows the trajectories of inputs associated to the weights in Fig.~\ref{fig:meanfield}(c) (corresponding to hidden variable $\mu=4$ and $14$). These two hidden variables are strongly activated at sites that are in contact in the tertiary structure of the protein (Fig.~\ref{fig:meanfield}(d)) and are consequentially relevant for its stability. While the logo of $w_{4}$ shows that the interaction between site 25 and its neighbors can be realized through electrostatic forces between charged amino acids~ $w_4$ tells that contacts between sites 5,6,11 and 22 can be realized through disulfide bonds between Cysteines (C). The dynamics of the projection $m^{14}$ (Fig.~\ref{fig:meanfield}(a)) explains how  global optimal paths exploit Cysteine-Cysteine interactions (not present in the initial and final sequences) in order to maintain the structural stability through  transient  mutations to C-C in the sites 5,6,11 of the protein. These  C-C  bonds  are then lost in the final configuration, as clearly seen by the decrease of the projection $m^{14}$.
%The exploration of favourable regions in the landscape is made possible by the slightly higher number of mutations between successive sequences in the former case than in the latter, see Fig.~\ref{fig:meanfield}(b). 
Along global paths, most of the intermediate mutational steps do not abruptly changes the order parameters, with the exception of the bump in the overlap $q$ at step $\sim$10, possibly related to the presence of preparatory mutations for the Cys-related transition in Fig.~\ref{fig:meanfield}(a). 

\begin{figure}
    \centering
    \includegraphics[width=\linewidth]{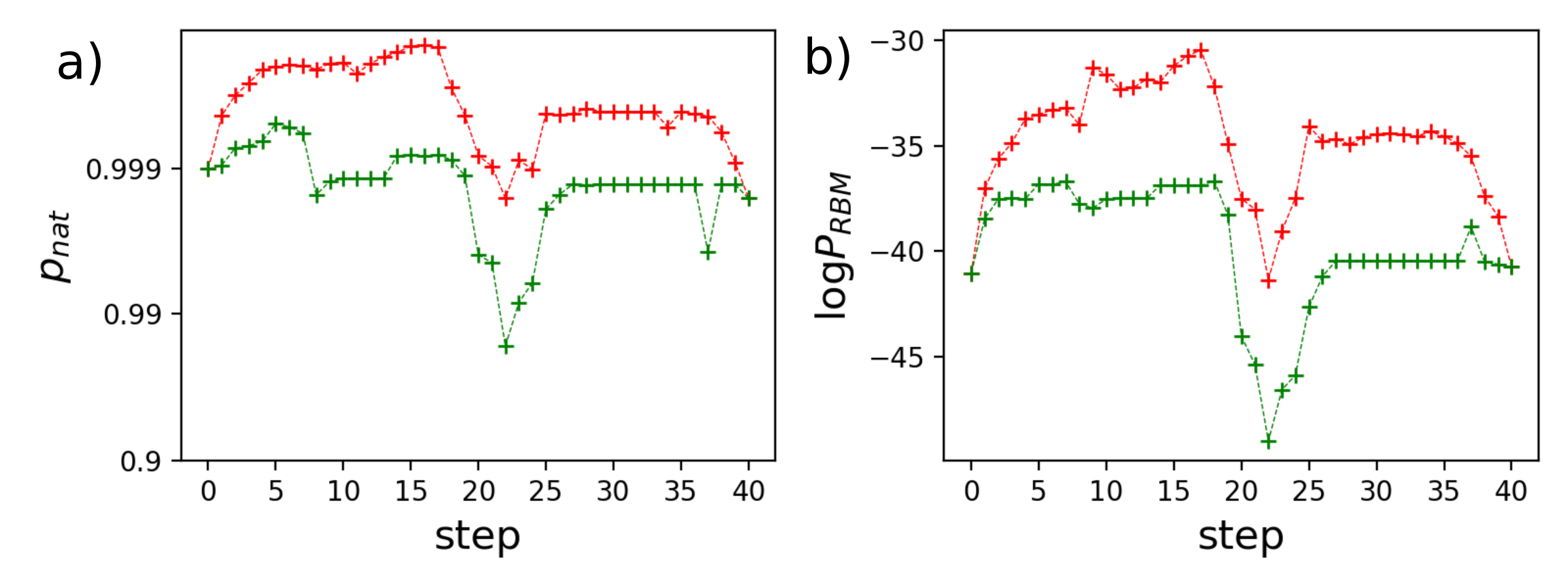}
    \caption{{\bf Average value of $p_{\text{nat}}$ and log-likelihood along the paths for lattice proteins} estimated from the mean-field global (red) and direct (green) solutions shown in Fig.~\ref{fig:meanfield}.}
    \label{fig:pnat}
\end{figure}

Using Eq.~\eqref{eq:freq} we can compute the amino acids frequencies at each site along the path and use this information to estimate the average log-likelihood and $p_{\text{nat}}$ at each step. To estimate the $p_{\text{nat}}$ we use this frequencies to build an independent site model that approximate the true marginal distribution of sequences, then we use this model to sample many sequences at a given step $t$ and compute the average $p_{\text{nat}}$ from this samples. The results shown in Fig.~\ref{fig:pnat} confirm very good values for the probabilities of intermediate sequences along the path, both for $p_{nat}$ (Fig.~\ref{fig:pnat}(a)) and for the model $P_{\text{RBM}}$ (Fig.~\ref{fig:pnat}(b)).
We also observe that sequences along the global paths have substantially higher probabilities than along direct paths for the values of $T$ and $\gamma$ considered.

\subsection{Application to  the WW domain}
\label{sec:prot2w}

We apply the above approach to RBM models learnt from sequence data of the WW family extracted from public database (PFAM id: PF00397)~\cite{WWstructure,WWspecificities}. WW is a small protein module with $\sim 30-40$ amino acids, able to specifically bind to peptidic ligands. In particular, we will study paths interpolating between two proteins known to have different binding activity:
\begin{itemize}
    \item[]$\vsec_{\text{start}}\,=\,$\texttt{LPAGWEMAKTSS-GQRYFLNHIDQTTTWQDP} \ ,
    \item[]$\vsec_{\text{end}}\,=\,$\texttt{LPKPWIVKISRSRNRPYFFNTETHESLWEPP}.
\end{itemize}
$\vsec_{\text{start}}$ was shown to have strong binding affinity to PPxY (x = any amino acid) motifs~\cite{espanel1999single} (called class I - WW domains), while $\vsec_{\text{start}}$ binds to pTP or pTS motifs (p=phosphorylated site)~\cite{russ2005natural} (called class IV - WW domains).

\subsubsection{Direct-to-global phase transition}
%LPAGWEMAKTSS-GQRYFLNHIDQTTTWQDP
%LPPGWEKRMSRSSGRVYYFNHITNASQWERP
%LPKPWIVKISRSRNRPYFFNTETHESLWEPP
\label{sec:dir_to_glob_WW}

\begin{figure}
    \centering
    \includegraphics[width=\linewidth]{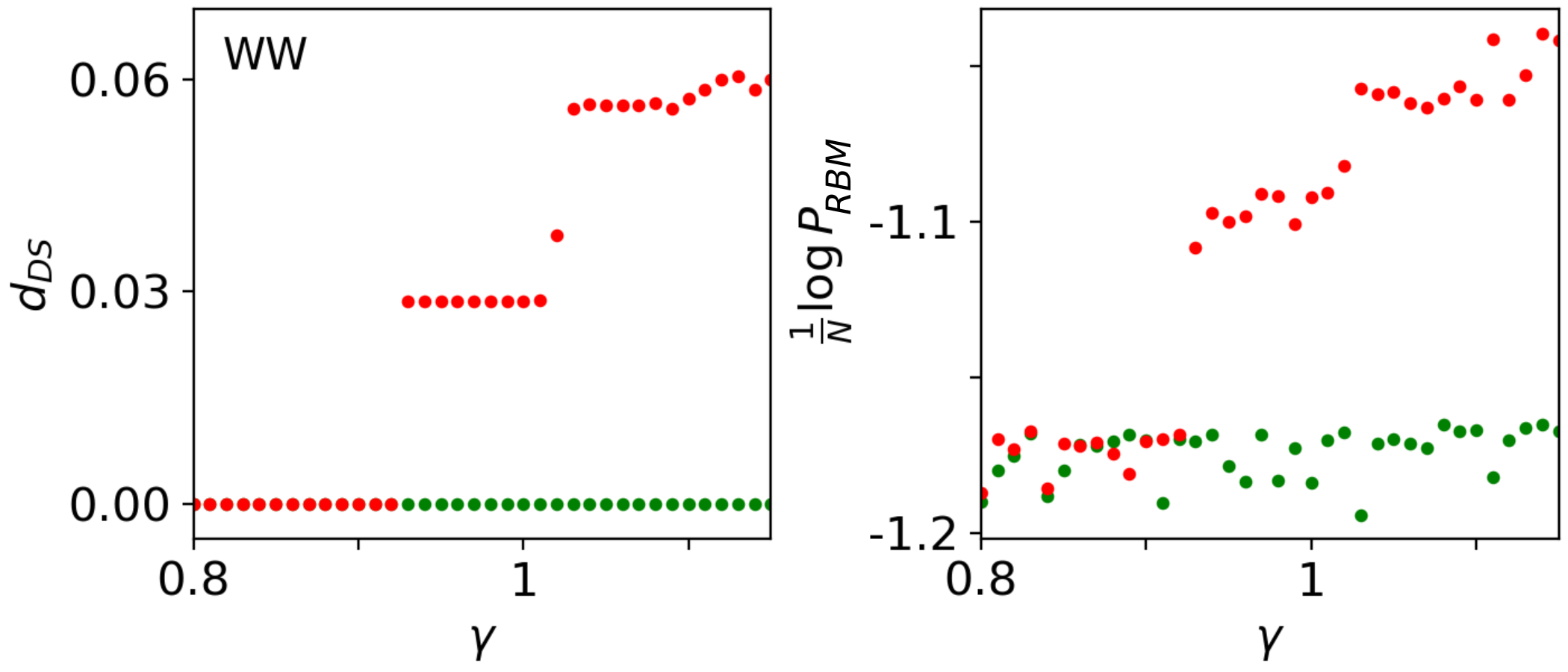}
    \caption{{\bf Direct-to-global phase transition in WW domain}. Mean-field estimates of $d_{\text{DS}}$ (\emph{Left}) and of $(\log P_{\text{RBM}})/N$ (\emph{Right}; red: global paths, green: direct) vs. $\gamma$ for mutational paths of the WW domain of length $T=10$. In all panels $\beta=3$. }
    \label{fig:WW_dirtoglob_pt}
\end{figure}

\begin{figure}
    \centering
    \includegraphics[width=\linewidth]{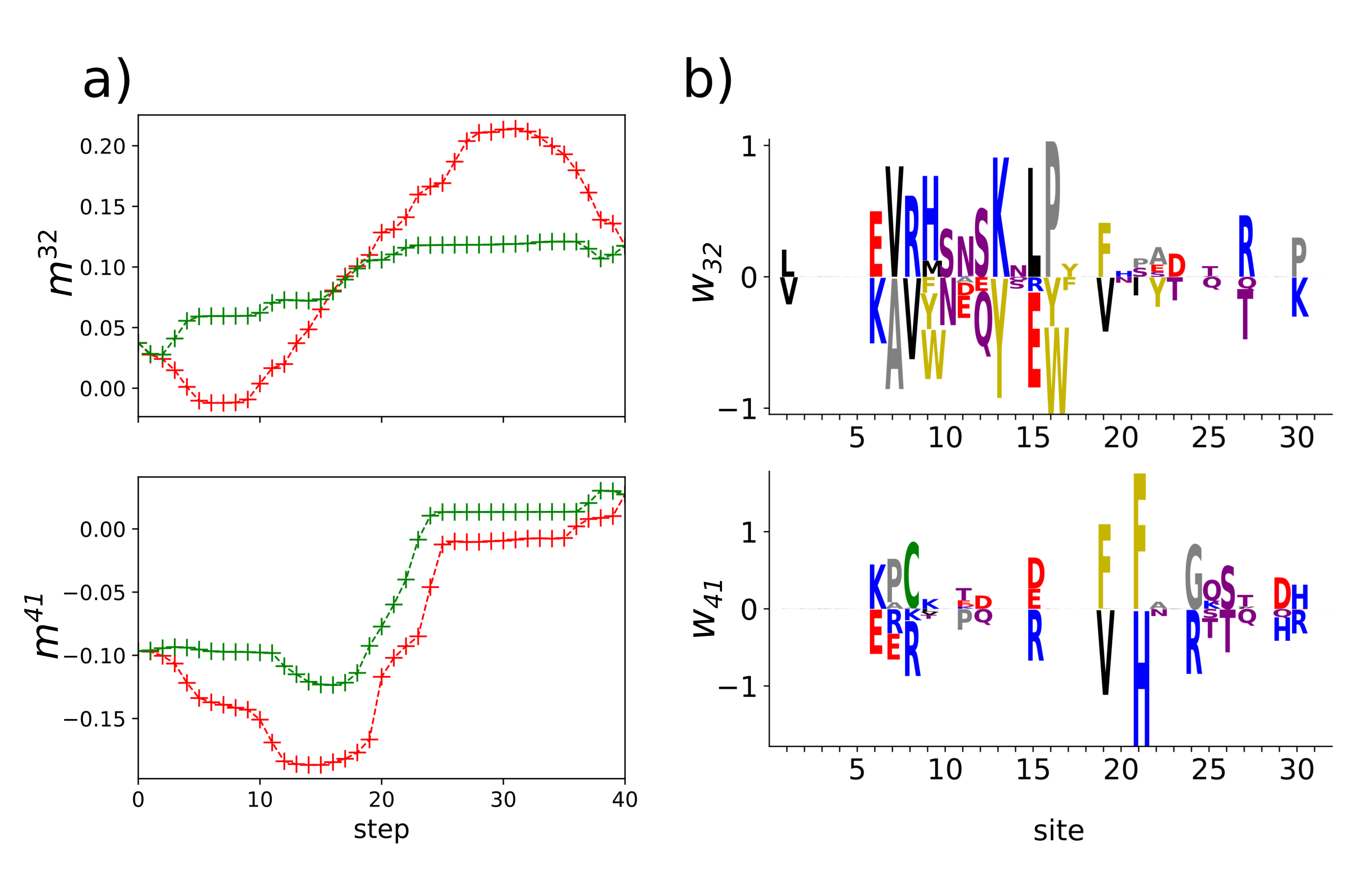}
    \caption{{\bf Direct and global transition path in WW domain}. (a) Values of two relevant inputs. Green and red paths correspond to direct and global solutions, respectively. (b) Logo of the weights associated to the inputs. Simulation parameter: $\gamma = 1.6$, $T=40$, $\beta= 3$. }
    \label{fig:WW_paths}
\end{figure}

Direct-to-global transitions are observed in mutational paths joining natural WW sequences, see Figure~\ref{fig:WW_dirtoglob_pt}. This figure shows in particular the presence of a cross-over, when the path length $T$ is kept fixed, between direct and global solution at a value of $\gamma\sim 0.92$ and another jump at $\gamma\sim1.3$, corresponding to the insertion of a novel mutation outside the direct space. 

\begin{figure}
    \centering
    \includegraphics[width=\linewidth]{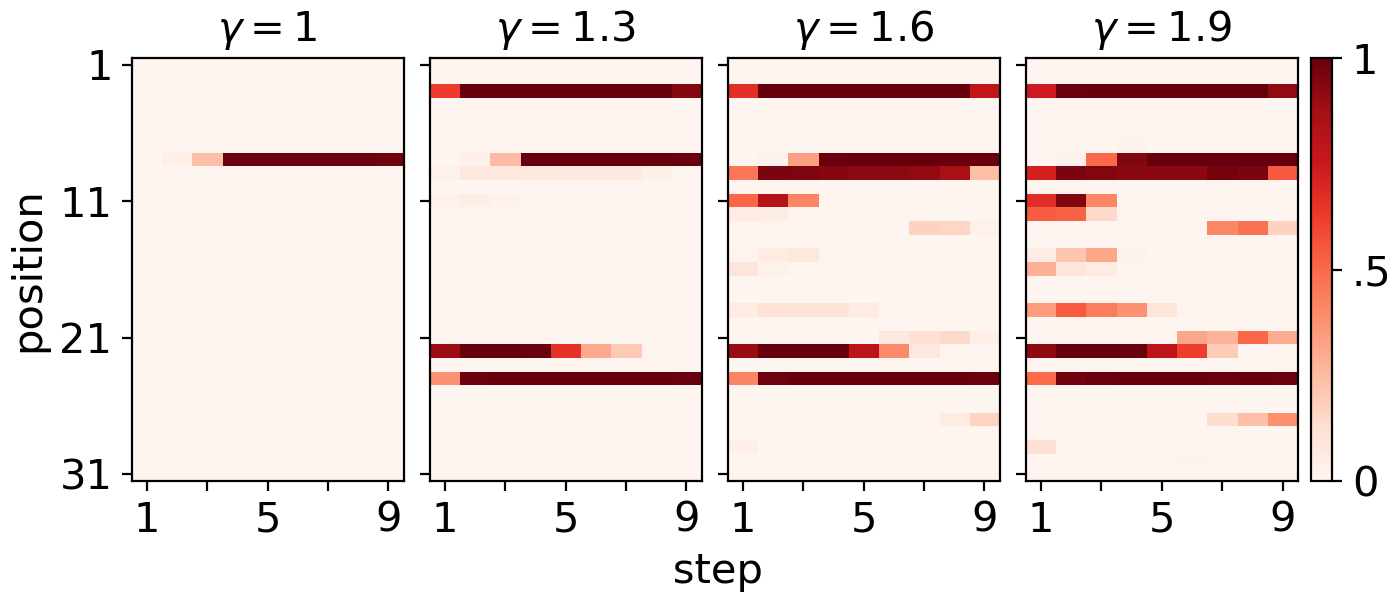}
    \caption{{\bf Probability of non-direct amino acids along direct paths} as a function of the step $t$ ($x$-axis) and of the sequence site $i$ ($y$-axis) for the WW domain. Results are shown for four values of $\gamma$, see panels. Parameter: $\beta=3$.  }
    \label{fig:WW_div_from_dir}
\end{figure}

\begin{figure}
    \centering
    \includegraphics[width=\linewidth]{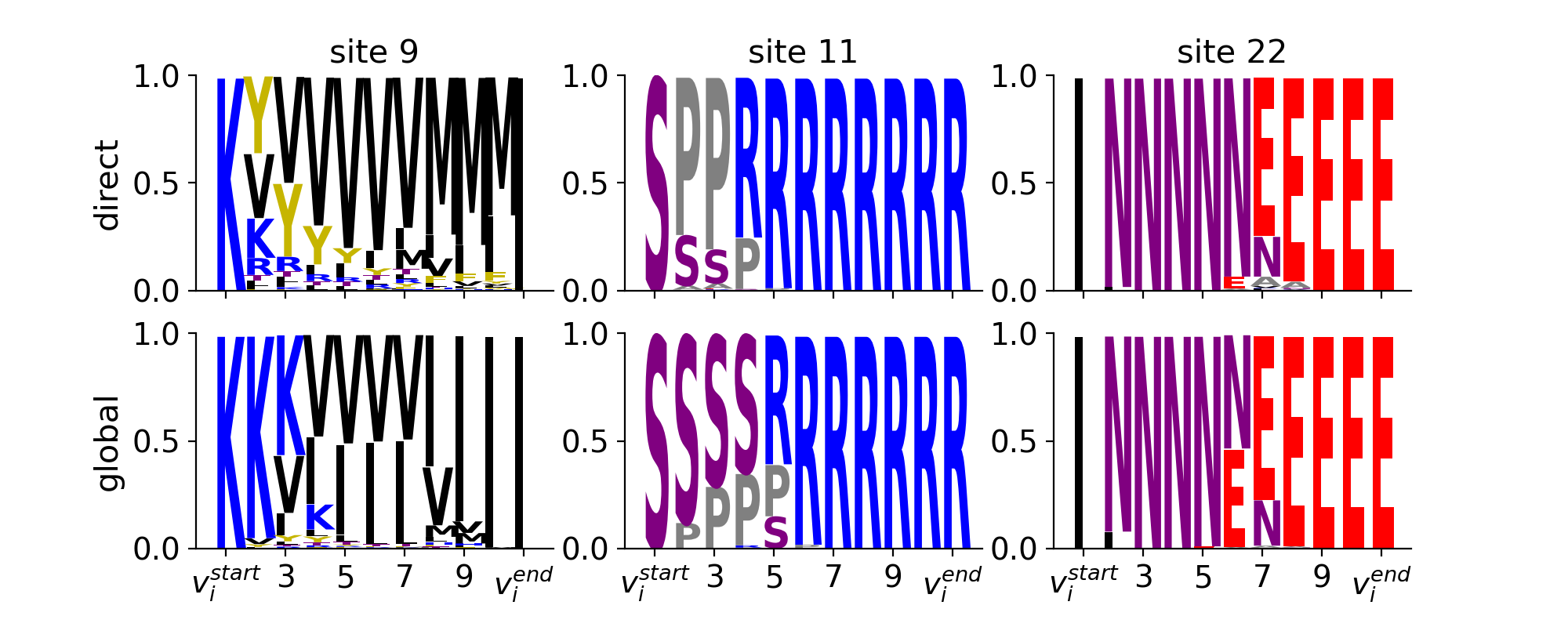}
    \caption{{\bf Logos of the amino-acid frequencies} at three arbitrarily chosen sites along a path of length $T=10$ joining two WW domains. (\emph{Top}) logos computed using the MF direct solution; the two amino acids allowed on each site in the direct subspace are the ones corresponding to $v^{\text{start}}$ and $v^{\text{end}}$. The other amino acids are candidate for mutations outside the direct space. (\emph{Bottom}) logos computed using the MF global solution. Here $\gamma=1.6$ and $\beta=3$.}
    \label{fig:logo_comparison}
\end{figure}

To further study the difference between direct and global solutions at different values of $\gamma$, we can compute what and where the first relevant mutations that push the solutions outside the direct space should be considered. Differently stated, given a direct path computed for certain value of length $T$ and potential stiffness $\gamma$, we would like to know what sites will be the first to mutate outside the direct space immediately after we release the constraint on the path to be direct (\emph{i.e.} we compute the mean-field solution only considering as accessible sites the ones present at the target sequences). To do so, we use Eq.~\eqref{eq:freq} to compute the frequencies of each amino acid $\langle \delta_{v_{it},a}\rangle$ in the global space (where the transfer matrix that defines $Z^{\text{1D}}_i$ is of size $21 \times 21$) around the direct solution. Then we compute the probability assigned to non-direct amino acids at some point by the direct mean-field solution, $p^{\text{out}}_{\text{DS}}(i,t)$, as:
\begin{equation}
    p^{\text{out}}_{\text{DS}}(i,t) = 1 - \langle \delta_{v_{it},v_{\text{start},i}}\rangle_{\#_{\text{dir}}}-\langle \delta_{v_{it},v_{\text{end},i}}\rangle_{\#_{\text{dir}}}\,.
\end{equation}
Results for different values of $\gamma$ are shown in Fig.~\ref{fig:WW_div_from_dir}. As expected for higher values of $\gamma$ the interaction potential $\pot_{\text{Cont}}$ becomes less stiff and allows the emergence of more mutations escaping the direct space. In the case of $\gamma=1$ the Cont potential is stiff enough to allow only one mutation outside the direct space. In particular this mutation appears in the middle of the path and stays until the very end (before returning to the final state at step $10$), showing that the path has to reach a proper region of the sequence space before engaging non-direct mutations. The difference between these global mutations computed on the direct solution and the global solution is shown in Fig.~\ref{fig:logo_comparison}, where we used Eq.~\eqref{eq:freq} to compute the frequencies of each amino acid. This approach can be useful to improve mutagenesis experiments by suggesting a minimal number of mutations outside the direct space that can already improve the quality of the intermediate sequences.\\

Differences between direct and global solutions in the case of WW domain can be observed in Fig.~\ref{fig:WW_paths}. Here we plot the values of two relevant inputs along both type of paths. The two weights have been chosen to between those that maximise the difference between direct and global solutions. In particular, we see that the projection along the weight $w_{32}$ for the direct solution remains almost constant compared to the global case. On the other hand, projection along the weight $w_{41}$ shows a switch in both cases, with global solutions showing a stronger activity.

\subsubsection{Entropy of doubly-anchored paths}
\label{sec:entropy_WW}

Our mean-field theory allows us to compute other quantities of interest, such as the number of relevant transition paths. Knowing the entropy of the distribution of paths would be useful for example to estimate  how rare the transition between two regions of the sequence space is.

From a practical point of view, despite the care brought in numerically solving  Eqs.~\eqref{eq:der_logZ_1d} a small disagreement between the left and right hand sides may subsist. As the number of order parameters scales proportionally to $T$ and $M$, these inaccuracies must be taken into account when estimating the entropy $\mathcal{S}_\text{path}$. To compute the latter we therefore estimate $f_{\text{path}}$ at different inverse temperatures $\beta$ and use the identity
\begin{equation}
   \mathcal{S}_\text{path}= - \frac{\mathrm{d} f_{\text{path}}}{\mathrm{d} (1/\beta)} \,.
   \label{eq:entropy_corr}
\end{equation}
This procedure gives a more precise estimate of the entropy than directly plugging the values of the order parameters in Eq.~(\ref{eq:entropy}).
In the case of RBM we obtain
\begin{multline}
        \mathcal{S}_{\text{path}} = - \frac{\beta}{N} \sum_{i,t} \langle g_{i,t}(a)\rangle + \frac{1}{N} \sum_i \log Z_i \\ 
       -\frac{\beta}{N}\left( \Gamma'(\mathbf{m}) \frac{\partial}{\partial \hat{\mathbf{m}}}  - \pot'(\mathbf{q}) \frac{\partial}{\partial \hat{\mathbf{q}}}\right) \sum_i \log Z_i\, .
\end{multline}

%Evaluated at the numerical solution $\mathbf{m}$, $\mathbf{q}$, $\hat{\mathbf{m}} = \beta \Gamma'(\mathbf{m})$ and $\hat{\mathbf{q}} = -\beta \pot'(\mathbf{q})$. 
Estimates of $\mathcal{S}_{\text{path}}$ in the Cont and Evo scenarios  are shown in Fig.~\ref{fig:evo_vs_cont}(a). The first important aspect to be noted regards the scaling of $\mathcal{S}_{\text{path}}$ with the path length $T$: while in the Evo scenario the entropy seems to grow linearly with $T$, we notice a slower growth $T$ in the Cont scenario. This behaviour can be understood in the following toy model. We consider a uniform (flat) landscape $P_{model}$, without constraint on the final sequence. In the Evo scenario, it is easy to show that each time step corresponds, on average, to a constant number of mutations whose value depends on $\mu$ and on $A$ only. Hence, the  entropy is approximately added the logarithm of this number at each step, and the total entropy will scale linearly with $T$. In the Cont scenario, the number of possible configurations at each step is bounded from above by the hard wall in $\pot_{\text{Cont}}$, defined by the  overlap $q_c = 1 - \gamma/T$. Considering that $\pot_{\text{Cont}}(q>q_c)\ll 1$, each sequence along a path will have on average $\rho N$  mutations with respect to the previous sequence, where $\rho=\gamma / T$ is the mutation probability per site. We then estimate the entropy of a binary variable (mutation or no mutation on each site) with probability $\rho$ is $-\rho\log \rho -(1-\rho)\log(1-\rho)\simeq \frac \gamma T \log T$ for large $T$. Hence, the total entropy (per site) of the paths of length $T$ is expected to scale as $\sim \log T$.

%Figure~\ref{fig:evo_vs_cont}(a) show different behaviours of the entropy in the Evo and Cont scenario, with the entropy of the constrained solution being higher than the unconstrained one in the Evo scenario (and the opposite happens with Cont). This arises from the fact that the unconstrained Evo solution in jumps directly from the initial configuration to the closest minimum of the energy landscape, while the constrained Evo solution has to interpolate distant target sequences. The presence of an hard wall potentials in the Cont scenario forbids both solutions to remain in the same configurations for long times and then jump directly to another distant point in sequence space. Hence, Cont solutions will explore many more different configurations making their entropy higher with respect to their Evo counterparts. Moreover, since the Cont constrained solution has to smoothly interpolate between distant region in such a way that the energy along the path is optimized, this makes the number of accessible paths lower with respect to the unconstrained solution.
%{\bf Eugenio: qualche commento sul valore di S? perche' e' piu basso con Evo che con Cont? Perche' e' piu basso nel caso unconstrained che nel caso constrained per Evo? Questo perche' probabilmente va a finire da un'altra parte pero' va spiegato.}

\subsubsection{Case of paths anchored at the origin}
\label{sec:escape}

\begin{figure}
    \centering
    \includegraphics[width=\linewidth]{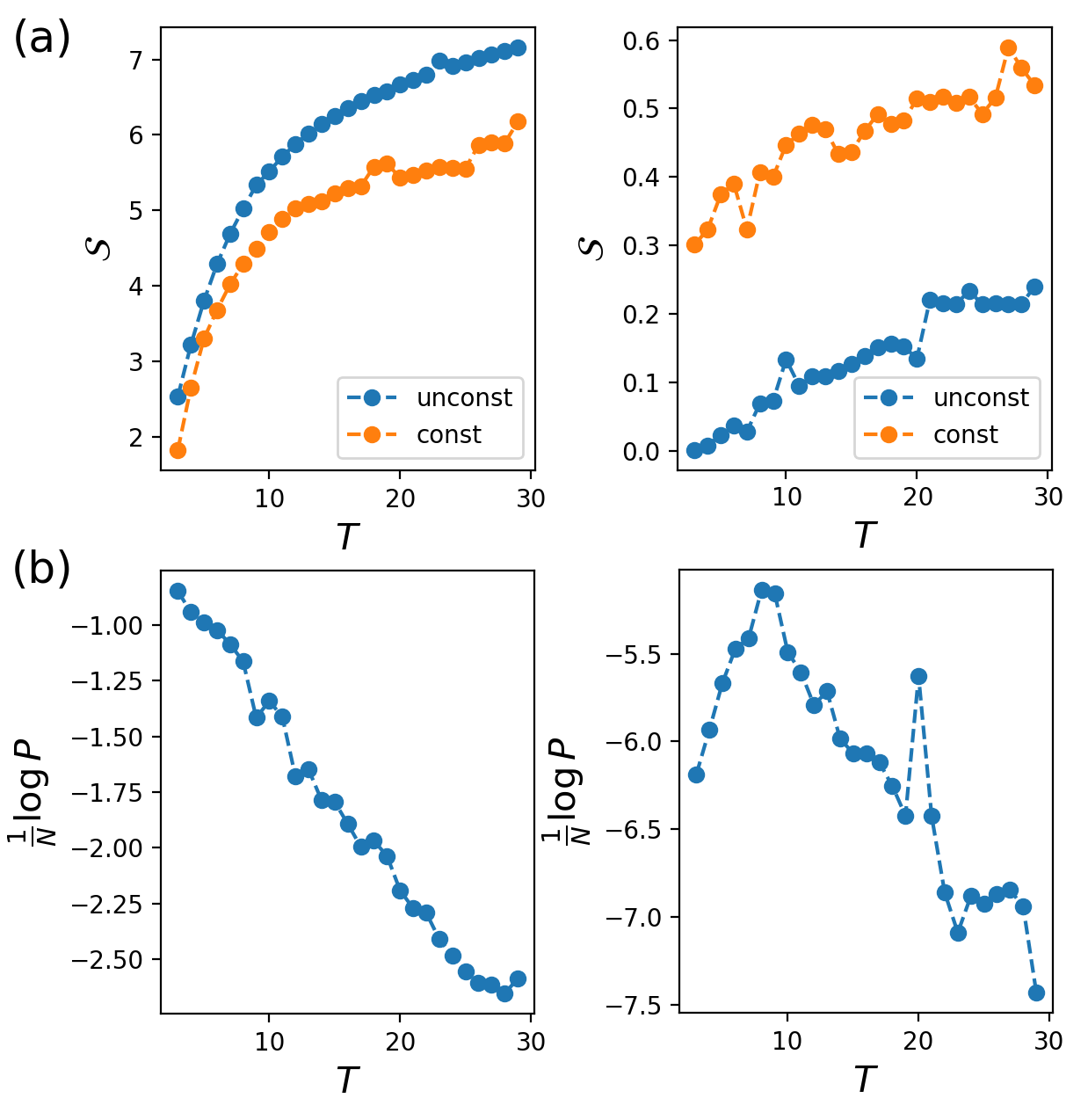}
    \caption{{\bf Entropies and probabilities of transition for the Cont (left) and Evo (right) potentials}. 
    (a) Entropy ${\cal S}_{\text{path}}$ of paths as a function of $T$. Results are shown for paths joining the two WW domain wild-type sequences (constrained) and paths anchored by the starting sequence and free at the other extremity (unconstrained). 
    (b) Probability of a transition path as a function of  $T$. 
    Parameters for Evo: $\mu=10^{-4}$, $\beta=1$; for Cont: $\gamma=3$, $\beta=1$. }
    \label{fig:evo_vs_cont}
\end{figure}

The partition function for paths in Eq.~\eqref{eq:Zpath} is computed on the ensemble of paths fixed at both ends to be equal to sequence $\mathbf{v}_{start}$ and $\mathbf{v}_{end}$. One can easily redo the computation when the last extremity is left free.  We show in Fig.~\ref{fig:evo_vs_cont}(a) the entropies of these partially unconstrained paths for the Cont and Evo potentials. 

In the Evo scenario the unconstrained solution shows lower entropy than the constrained one, while it as a higher entropy in the Cont scenario as intuitively expected. 
%This arises from the fact that the unconstrained Evo solution in jumps directly from the initial configuration to the closest minimum of the energy landscape, while the constrained Evo solution has to interpolate distant target sequences.
This apparently surprising finding can be explained as follows. For the constrained, doubly- anchored paths $\mathbf{v}_\text{end}$ has relatively high energy (see Fig.~\ref{fig:pnat}(b)), and many paths connect this last sequence to $\mathbf{v}_\text{start}$. Conversely, in the unconstrained case, paths are attracted to a lower free-energy minimum, and there are fewer paths connecting the initial configuration to this final region. 
The presence of a hard wall in the Cont scenario forbids both solutions to remain in the same configurations for long times and to then jump directly to another distant point in sequence space. Hence, Cont solutions will explore many more different configurations making their entropy higher with respect to their Evo counterparts. Moreover, since the constrained solution in the Cont case has to smoothly interpolate between distant regions in such a way that the energy along the path is optimized, this makes the number of accessible paths lower than in the unconstrained solution.%This can be explained by the fact that the unconstrained solution in the Evo scenario flows quickly into the closest minimum of the free energy since it is not limited in the number of mutations contrary to the Cont case) between each step.

Our mean-field formalism allows us to compute the probability to go from $\mathbf{v}_{start}$ to $\mathbf{v}_{end}$ in $T$ dynamical steps, see \cite{mauri2023}. 
%\begin{multline}
%    P(\mathbf{v}_{start}\to\mathbf{v}_{end}|T) = \frac{\sum_\pathV^{\text{const.}}e^{-N\mathcal{E}(\pathV;\mathbf{v}_{start},\mathbf{v}_{end})}}{\sum_\pathV^{\text{unconst.}}e^{-N\mathcal{E}(\pathV;\mathbf{v}_{start})}} \\ 
%    \underset{{\scriptscriptstyle N\gg1}}{\sim}\frac{e^{-Nf_{\text{path}}^{\text{const.}}(\mathbf{v}_{start},\mathbf{v}_{end}|T)}}{e^{-Nf_{\text{path}}^{\text{unconst.}}(\mathbf{v}_{start}|T)}}\, .
    \label{eq:tran_prob}
%\end{multline}
This probability acquires an evolutionary interpretation in the case of the Evo potential. It estimates the probability to join the two sequences in $T$ steps consisting of mutations at rate $\mu$ (per step) combined with selection with probability $P_{model}$. We show in Figure~\ref{fig:evo_vs_cont}(b) the transition probabilities for the Cont and Evo scenarios. The Evo scenario shows an optimal length $T^*$ for which the probability is maximised, while, in the Cont scenario,  the transition probability decreases linearly  with $T$. This may be explained from the fact that the Evo potential emulates a mutational dynamics in which $T^*$ plays the role of an evolutionary distance between the two edge sequences. On the contrary the emergence of this optimal $T^*$ is forbidden in Cont scenario by the stiffness of $\pot_{\text{Cont}}$, which increases with $T$.

\begin{figure}
    \centering
    \includegraphics[width=\linewidth]{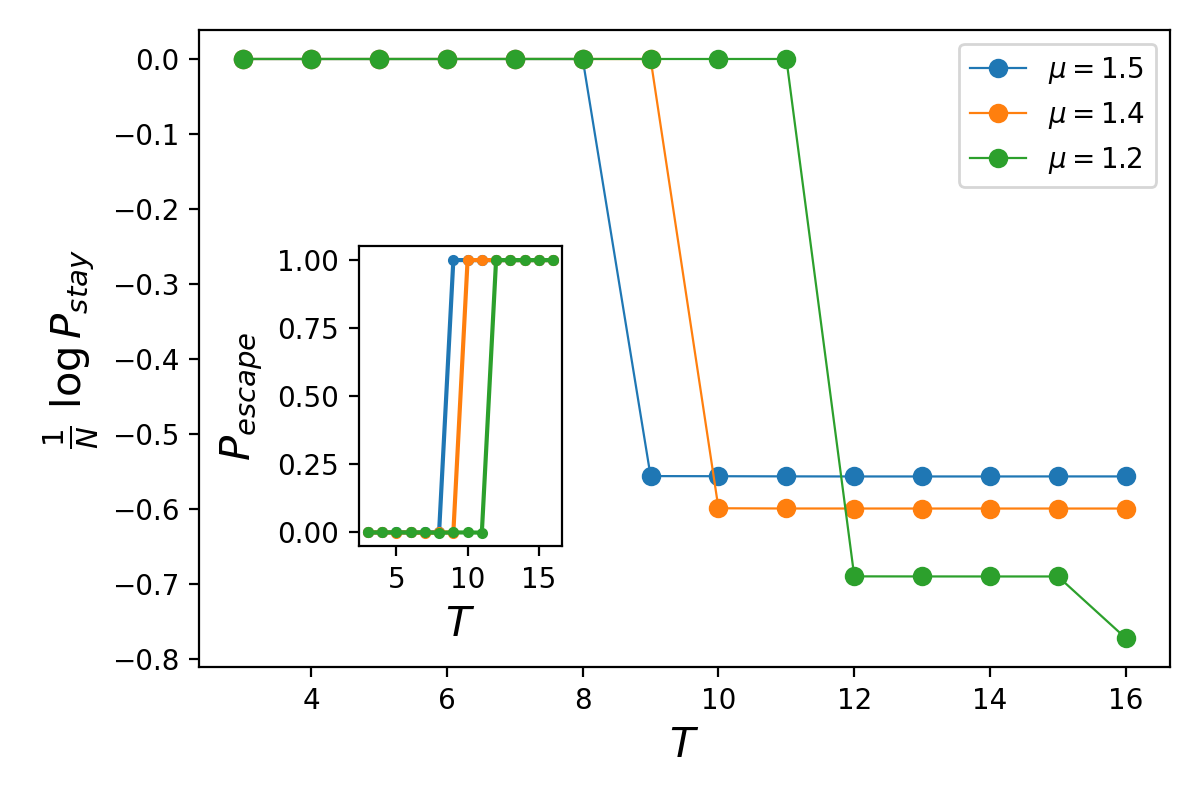}
    \caption{{\bf Probability of remaining in (main panel, log. scale along $y$-axis) and of escaping from (inset) the neighborhood of $\mathbf{v}_{start}$} for the WW domain, computed using Eq.\eqref{eq:escape_prob_HP} in the Evo scenario ($\beta=1$). Three different values of the mutation rate $\mu$ are considered.}
    \label{fig:escape}
\end{figure}

Furthermore, the framework above allows us to compute the probability of remaining in the minimum of the free-energy landscape corresponding to the starting sequence towards some region $\mathcal{R}$ of the sequence space in $T$ steps. We define  $P_{\text{stay}}(\mathcal{R}|T)$ as in
Eq.~\ref{eq:escape_prob_HP}. In Fig.~\ref{fig:escape} we plot the probability of remaining in the region associated to $\mathbf{v}_\text{start}$  for the WW domain energy landscape and in the Evo scenario. For different values of $\mu$ we are able to estimate at which time an evolving configuration is supposed to escape from the minimum. We observe the existence of a trade-off between the time and the probability of sojourn in the starting region depending on the value of $\mu$.

\section{Conclusion}
\label{sec:conclusion}

In the present work, we have focused on the study of transition paths in Potts-like energy landscapes in high dimension $N$. These paths can  be anchored at the initial and final extremity, or at the origin only.  Paths explore the energy landscape under conflicting constraints. First, contiguous configurations along the path should differ little from each other, in a way controlled by an elastic potential.  Second, intermediate configurations should have low energies, or, equivalently, high probabilities in the landscape. 

We have considered two kinds of elastic potentials: the first one, referred to as Cont,  ensures a smooth interpolation between sequences along the path and avoid `jumps' between configurations. 
The second potential, called Evo is inspired by evolutionary biology  {\em i.e.} it mimics random mutations at a constant rate $\mu$~\cite{leuthausser1987statistical}, while the energy landscape plays the role of the selective pressure driving the evolution. To avoid local maxima in the landscape successive intermediate along Evo paths may occasionally differ by more than the average number of mutations, $\mu N$. 

Using mean-field theory, we have computed  the typical properties of Evo and Cont paths in two  contexts. The first one, called direct ($dir$)  interpolates between two edge sequences, assigning on each site along the path one of the two active states present at the fixed edges. If the Hamming distance between the two extremities of the path is $D$, there are $2^D$ distinct direct intermediate sequences. The second one, called global ($glob$), may introduce novel mutations along the path compared to the target sequences, allowing for a deeper exploration of the energy landscape. While global paths can find better (\emph{i.e.} with lower energy) intermediate sequences, they are associated to higher elastic potential energy  due to the fact that global paths are in general longer (in terms of total number of mutations) than direct paths. Whether the subspace of direct paths is statistically dominant in the set of all possible global paths depends on their length and on their flexibility, controlled by the elastic potential. 

In the Cont case, we have unveiled the existence of  a direct-to-global phase diagram controlled by the stiffness of the interacting potential and the total number of steps of the path, together with the inner structure of the energy landscape. We have analytically described this phase diagram for the so-called Hopfield-Potts model, with only two interaction patterns with projections outside the direct subspace of controlable amplitude. We have analytically located the direct-to-global phase transition in a low temperature/high length regime as a trade-off between long, flexible paths with low energy intermediate configurations and short, stiff paths minimizing the number of mutations to go from one sequence to another. In this low temperature regime, the direct-to-global transition is essentially not affected by the number $A$ of Potts states (colors). Conversely, in the high temperature regime, that is, if the fluctuations of the energy are smaller than, or comparable to the inverse of the path length, paths tend to be global due to thermal fluctuations and the entropy of the system will depend on the total number of accessible state $A$ per site.

This direct-to-global phase transition takes place due to the conditioning on the final extremity of the paths. While evolutionary paths are generally not constrained in this way, there exist relevant situations in which conditioning is important. For instance, consider a directed evolution experiment starting from a wild-type sequence (of DNA, RNA, protein). Samples of the pool of sequences are retained at each round of selections/mutations. After several rounds, a sequence is obtained, and one asks for the possible transition paths that led to this outcome from the wild type. This well-posed question can be addressed with the methods proposed in this work, and confronted to sequences sampled at intermediate rounds. In addition, irrespective of conditioning at the end of the path,  we have shown that the direct-to-global transition is intimately related to the presence of attractive region in the energy/fitness landscape (Fig.~\ref{fig:HPsketch}).

From a statistical mechanics point of view, the mean-field approach followed here computes transition paths for a given  realization of the quenched disorder. This is made possible by the fact that, formally, the number $M$ of patterns in the Hopfield-Potts model (or of hidden units in the RBM) is finite as $N\to\infty$. We plan in future  to extend our approach with $M$ scaling linearly with $N$. A possible application, in the case of RBM, would be the so-called compositional phase of \cite{tubiana2017emergence}, where each data configuration activate a finite number of hidden units. In particular, in this scenario we aim to describe the free energy of the system as only a finite number of patterns are active, while the other acts as a white noise.

Last of all, we have tested our method for computing transition path on to data-driven models of natural proteins, extending the previous work  \cite{mauri2023} by showing how we could compute different quantities of interest, such as the entropy, \emph{i.e.} the number of relevant transition paths, the transition probability between two sequences, and the escape probability from confined regions of the sequence space.
Future work are definitely needed to improve our approach, \emph{e.g.} by considering finite-$N$ fluctuations around the mean-field theory solution. From a biological point of view, understanding the shape and the connectivity of the protein fitness landscape, and its entropy is of fundamental importance in the field of natural evolutionary processes and also for directed evolution experiments. The motivation is here not only theoretical but also practical, \emph{e.g.} to gain intuition on how many random sequences can evolve a given functionality under selective pressure. As stressed out in \cite{mauri2023}  inferring the optimal paths  and its optimal length ($T$) with the Evo potential is an extension of the reconstruction of phylogenetic trees and of the optimal evolutionary distance between two ancestral sequences for epistatic fitness landscapes, inferred from data.   Finally,  better characterizing transition paths could help predict escaping mutations,\emph{e.g.} allowing a virus to escape from the control of the immune system, and is therefore of primary importance in the development of effective drugs or vaccines.

\vskip .3cm
\noindent
{\bf Acknowledgments.} This work was supported by the ANR-19 Decrypted
CE30-0021-01 project. E.M. is funded by a ICFP Labex
fellowship of the Physics Department at ENS.
\typeout{}
\bibliography{citations,citations2}

\end{document}